\documentclass[aps,pra,reprint,superscriptaddress,groupedaddress,noshowpacs]{revtex4-2}  
\usepackage{graphicx} 
\usepackage{amsmath}
\usepackage[colorlinks]{hyperref} 
\usepackage[caption=false]{subfig}
\usepackage[section]{placeins}
\usepackage{xcolor}

\begin{document}
\title{Bogoliubov-de Gennes theory of the `snake' instability of gray solitons in higher dimensions}
\affiliation{University of Kaiserslautern, Germany}
\author{Alexej Gaidoukov} 
\author{James R. Anglin}\affiliation{University of Kaiserslautern, Germany}
\begin{abstract}
Gray solitons are a one-parameter family of solutions to the one-dimensional non-linear Schr\"odinger equation (NLSE) with positive cubic nonlinearity, as found in repulsively interacting dilute Bose-Einstein condensates or electromagnetic waves in the visible spectrum in waveguides described by Gross-Pitaevskii mean field theory. In two dimensions these solutions to the NLSE appear as a line or plane of depressed condensate density or light intensity, but numerical solutions show that this line is dynamically unstable to `snaking': the initially straight line of density or intensity minimum undulates with exponentially growing amplitude. To assist future studies of quantum mechanical instability beyond mean field theory, we here pursue an approximate analytical description of the snake instability within Bogoliubov-de Gennes perturbation theory. Within this linear approximation the two-dimensional result applies trivially to three dimensions as well, describing buckling modes of the low-density plane. We extend the analytical results of Kuznetsov and Turitsyn [Sov. Phys. JETP \textbf{67}, 1583 (1988)] to shorter wavelengths of the `snake' modulation and show to what extent the snake mode can be described accurately as a parametric instability, in which the position and grayness parameter of the initial soliton simply become dependent on the transverse dimension(s). We find that the parametric picture remains accurate up to second order in the snaking wave number, if the snaking soliton is also dressed by an outward-propagating sound wave, but that beyond second order in the snaking wave number the parametric description breaks down.  
\end{abstract}

\maketitle

\section{Introduction}
\subsection{A soliton instability}
Among the many reasons for interest in solitons is their appearance in quantum many-body systems, for example as solutions to the Gross-Pitaevskii (GP) mean field theory of a dilute Bose-Einstein (quasi-)condensate \cite{Sengstock,canary}. In one-dimensional scenarios with repulsively interacting condensates, so-called \emph{dark} or \emph{gray solitons} are robust objects, but in two or three dimensions they exhibit the `snake' instability \cite{First_Snake,Zakharov,Snake2,Huang,Feder,Brand,VRN}. 
Within the classical mean-field approximation and in (effectively) two dimensions the snake instability can easily be followed numerically on a desktop computer; an example is shown in Fig.~\ref{fig:GPev}. 

The snake instability has been examined numerically in many papers and has also received previous analytical treatments. Several of these latter, however, have focused on the Kadomtsev-Petviashvili equation \cite{First_Snake,Snake_KP,KP2000}, to which the GP equation reduces in the `shallow soliton' limit where the amplitude of the soliton is small compared to the background density. Here instead we will present detailed analytical results for gray solitons of arbitrary depth, under the full GP nonlinear Schr\"odinger equation, using perturbation theory in the snake-mode wave number. Our explicit derivations will confirm formulas for the snake instability growth rate that have been presented previously without derivations that were valid for all soliton depths. Our main purpose, however, is to show exactly what happens in the snake instability, and not only how fast it happens, by deriving explicit expressions for the growing perturbations to the space-dependent order parameter. These explicit results for the mode functions are necessary ingredients for future investigation of the snake instability beyond mean field theory, taking thermal and quantum fluctuations into account.

\subsection{Towards the quantum instability}
\begin{figure*}[htbp]
	\centering
	\includegraphics[width=0.7\textwidth,trim={1.9cm 2.9cm 7cm 4cm},clip]{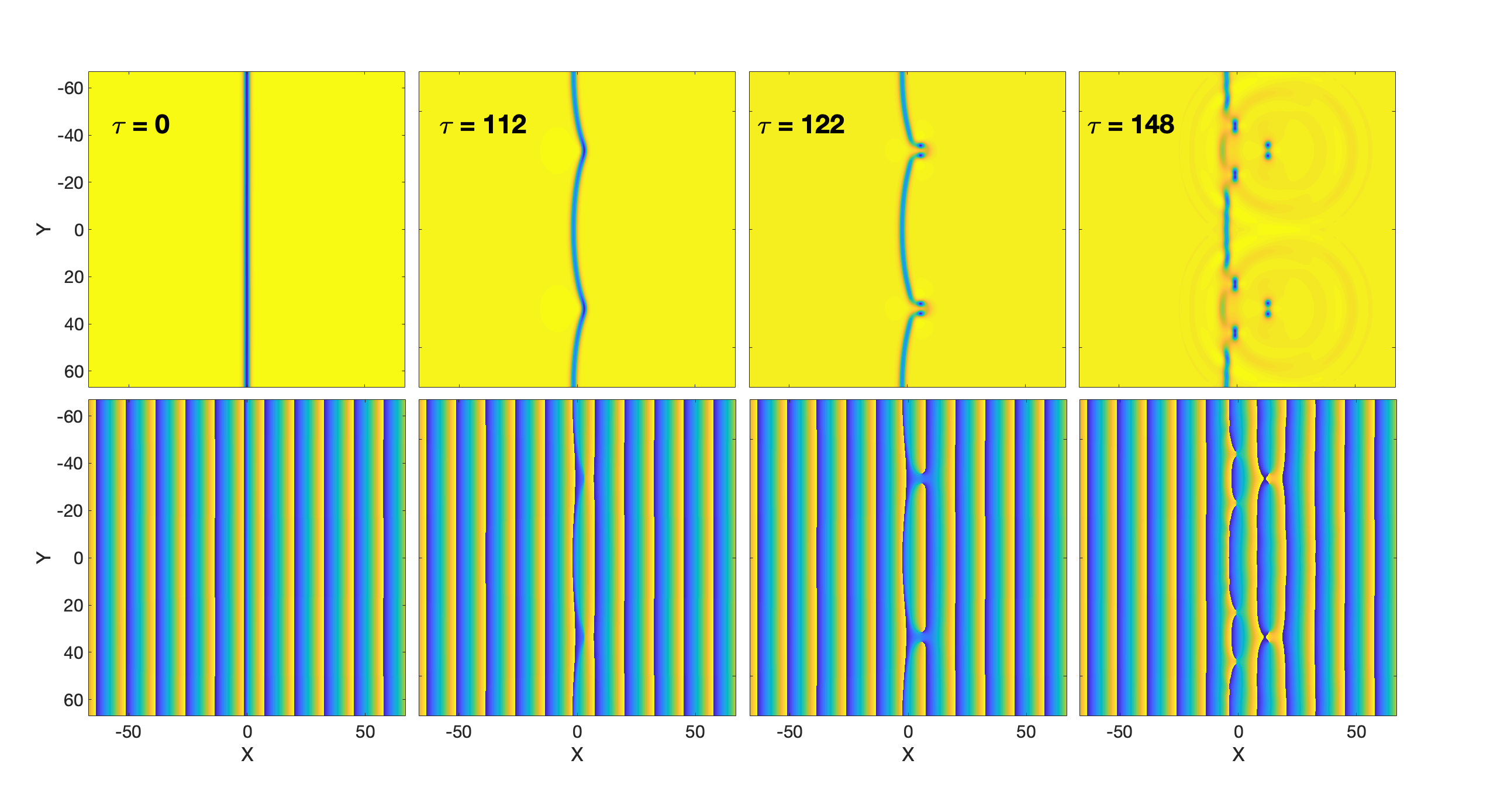}
	\caption{\textbf{The Snake Mode.} Gross-Pitaevskii evolution of condensate order parameter modulus $|\Psi|$ (upper plots) and phase $\arg(\Psi)$ (lower plots) at different times (indicated in units of the inverse chemical potential $\hbar/\mu$ by the dimensionless parameter $\tau$) from an initial gray soliton with a perturbation that is initially too small to be seen. The well-known `snake instability' makes the initial density trough undulate like a crawling snake. Ultimately quantized vortices and anti-vortices appear, as seen in the lower right frame, where there are four points around which the phase sweeps through the full color range representing 0 to $2\pi$, respectively being a singularity of the phase. The spatial axes $x$ and $y$ are in units of the condensate healing length, as explained in Section II.}
	\label{fig:GPev}
\end{figure*}

Understanding the gray soliton snake instability within non-equilibrium quantum statistical mechanics is an interesting problem because, as we see in Fig.~\ref{fig:GPev}, the snake instability leads to the formation of quantized vortices. It has been confirmed experimentally that this occurs with real quantum gases \cite{SnakeExp}, not just in GP mean field theory. Beyond the mean field approximation, therefore, the snake instability offers an interesting opportunity to study quantum many-body phenomena, with comparison between theory and experiment, because the vortices that emerge from the decay of the soliton will be experimentally observable signals of quantum fluctuations that have effectively been amplified by the instability. Theoretical understanding of such mesoscopic quantum effects remains challenging, however. As a starting point for later quantum investigations, it will be useful to have a simple but accurate analytical description of the gray soliton snake instability, at least in its earliest stages of growth.

In these early stages the snake instability can be described within the Bogoliubov-de Gennes (BdG) linearization of the GP nonlinear Schr\"odinger equation. Since the initial background soliton is a one-dimensional structure extended to higher dimensions, the background has translation symmetry in transverse directions. Linearized excitation modes around the soliton can therefore have definite wave numbers in these directions, either $k=k_\perp$ for a line soliton in two dimensions, or $k=|\mathbf{k}_\perp|$ for a plane soliton in three dimensions; without loss of generality for the linearized problem we consider two dimensional scenarios from now on. Analytical BdG solutions were obtained by Kuznetsov and Turitsyn in the limit of small $k$, describing long-wavelength snake instabilities \cite{Kuznetsov}. These analytical solutions are moreover of promisingly simple form: as we will review below, they suggest that the snake instability might be accurately described as a parametric instability, such that its functional form remains within the gray soliton family but its position, phase, and other parameters shift in a way that depends sinusoidally on $y$ and exponentially on $t$. If the snake instability were even approximately this simple, it would be convenient for the analytical theory of the snake instability as a \textit{quantum} dynamical instability, because one could then hope to describe the quantum snake instability in terms of a quantized collective coordinate $\hat{x}_0(y,t)$ of the gray soliton---a considerable simplification from having to consider the full field operator $\hat{\psi}(x,y,t)$ of the entire quantum gas.\\

An example of how the parametric nature of the snake instability could be exploited has recently been provided by  \cite{KWCGF2017}, which pursues the ``Landau dynamics'' approach of modifying the one-dimensional gray soliton energy into a functional of $t$- and $y$-dependent soliton position. This approach in \cite{KWCGF2017} is somewhat phenomenological; the effective Lagrangian is motivated by a general appeal to adiabatic invariance without any explicit Ansatz for the two-dimensional order parameter being defined. Moreover the results of Ref.~\cite{KWCGF2017} indicate a need for deriving the effective Lagrangian for the collective coordinate more rigorously, for they yield a linearized instability growth rate for \emph{all} wavelengths which agrees with the \emph{long}-wavelength result in \cite{Kuznetsov}. This agreement is impressive for a first approximation, but problematic for precise comparison with experiments, because the long-wavelength limit of the snake instability is not actually representative of all wavelengths---and it is the shorter wavelengths which will be seen in experiments.

\begin{figure}[htbp]
	\centering
	\includegraphics[width=0.45\textwidth,trim={4.5cm 0cm 8.5cm 3.5cm},clip]{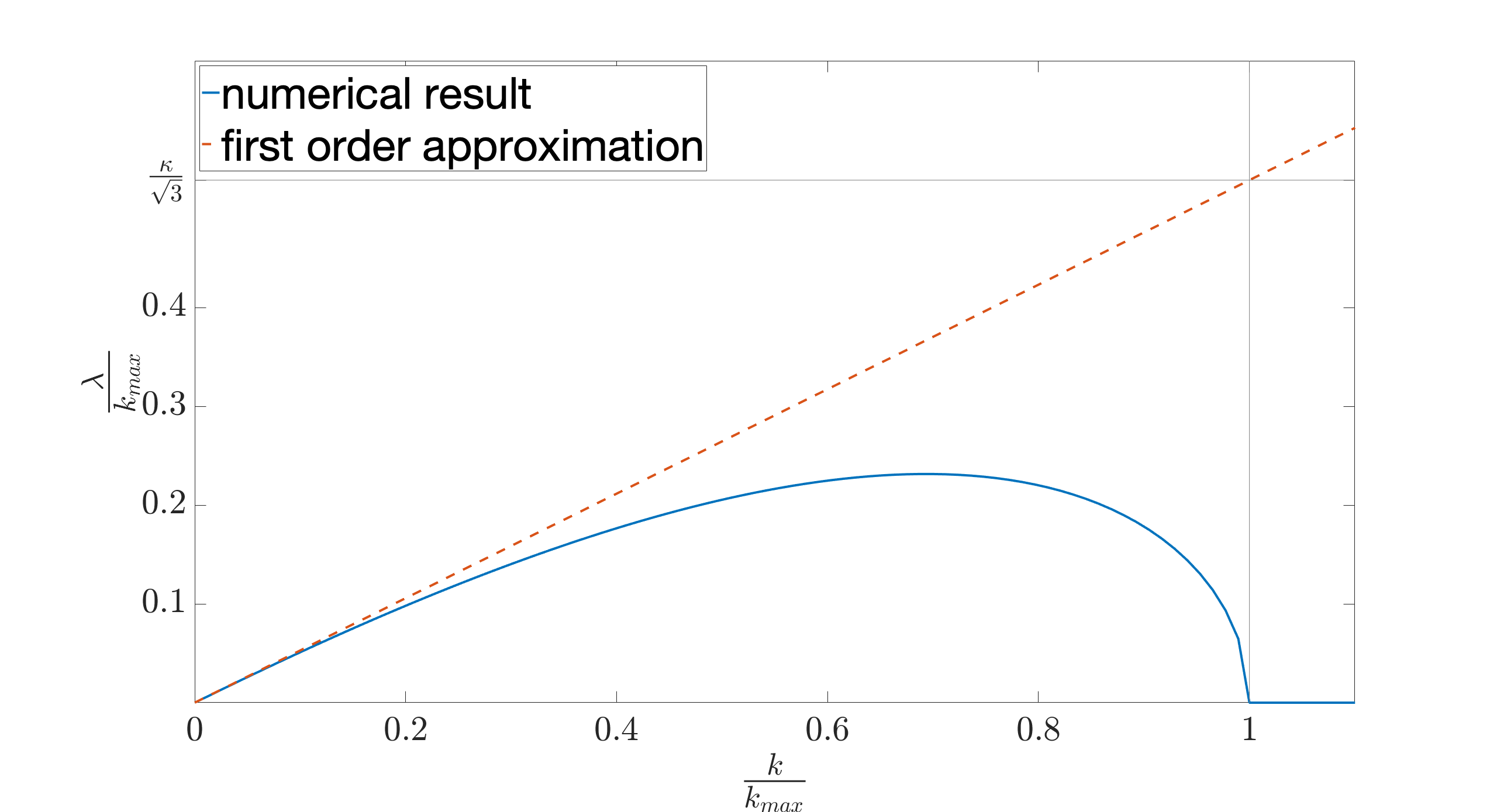}
	\caption{\textbf{Growth rate $\lambda/k_{\text{max}}$ vs $k/k_{\text{max}}$ plot.} The plot shows the numerical calculation (blue line) and the linear approximation (red line, having slope $\frac{\kappa}{\sqrt{3}}$) of $\lambda(k)$ normalized by $k_{\text{max}}$. The linear approximation agrees with the numerical result for small $k$ but deviates quickly for larger $k$.}
	\label{fig:k_l}
\end{figure}

\subsection{Beyond small $k$}
The reason why shorter wavelength snaking must be expected in experiments is that the growth rate for the long-wavelength instabilities, as found by Kuznetsov and Turitsyn, is \emph{slow} ($\propto k$). Whether the snake instability is initiated from thermal or quantum fluctuations, or from perturbations in the preparation of the initial soliton, unstable modes of all possible $k$ must be expected, and the most rapidly growing ones will typically be those that are actually seen. As Fig.~\ref{fig:k_l} shows, the most rapidly growing modes are of larger $k$, beyond the small-$k$ limit analyzed in \cite{Kuznetsov}. If we wish to understand finite-$k$ snake modes analytically, therefore, the results of \cite{Kuznetsov} must be extended to higher $k$. 

It is furthermore worth noting that Fig.~\ref{fig:k_l} also confirms another result of \cite{Kuznetsov}: the range of unstable $k$ is finite, extending only up to a maximal $k_{\text{max}}$, beyond which snaking perturbations no longer grow. \cite{Kuznetsov,PSK1995}. This fact provides additional motivation to extend the results in \cite{Kuznetsov} to higher $k$, inasmuch as there is only a finite domain of $k$ which needs to be covered, and it may be possible to reach the range of most rapidly growing modes with only a bit more work.

In this paper we therefore apply the method of matched asymptotics \cite{MatchAsym} to extend the results of Kuznetsov and Turitsyn, for BdG modes in a two-dimensional gray soliton background, to higher order in modulational wave number $k$. Our main goal and result is computing the mode functions explicitly, to find what actually happens to the condensate order parameter in space, as the instability grows in time, up to order $k^2$. In the course of computing the spatial mode functions for the instability to these higher orders, however, we also find the growth rate to order $k^{3}$. Although this is not our main goal in this paper, we will compare it and its derivation to previous results \cite{KP2000,KP2008}, and to numerical rates for all $k$, in order to assess how well our second-order mode functions are likely to represent snake instabilities seen in experiments.

\subsection{What we will find}
From our results for snake mode spatial dependence, our conclusions for future studies of quantum and thermal fluctuations will be a mixture of good and bad news. On the one hand we find that the parametric nature of the snake instability (dressed by a sound wave) persists to order $k^2$, at least to within a good approximation, and that an approximation which stops at order $k^2$ should still be able to come fairly close to the most rapidly growing unstable wavelengths. On the other hand we find that beyond order $k^2$ the snake instability becomes non-parametric, in the sense that BdG mode functions for the snake instability no longer correspond to simple modulations of the gray soliton background wave function, but instead begin to involve more obscure special functions (dilogarithms at order $k^3$). We conclude that for future quantum mechanical studies based on analytical BdG solutions, the second-order mode functions will provide some significant improvement in accuracy for more rapidly growing unstable modes, but that if higher accuracy is required for the most rapidly growing modes, there is unfortunately no practical advantage in using analytical solutions, because beyond order $k^{2}$ they will be just as complicated and opaque as purely numerical solutions.

Our presentation is structured as follows. In Section~II we review the appropriate GP and BdG equations for our scenario, including the particular steady-state solution that represents a gray soliton extended uniformly into two dimensions. Then in Section III we will use the multiple-scale analytical method of matched asymptotics to derive explicit approximate forms for the imaginary-frequency BdG normal modes that represent the snake instability, for cases where the wave number $k$ of the snake perturbation is small. These results will reproduce those of Ref.~\cite{Kuznetsov}. Section IV will present our further advances, gained by applying the methods of Section III to higher orders in $k$ perturbatively; Section IV will be shorter than Section III because it will only present results, with their rather lengthy derivations reserved for Appendix A. We will then conclude in Section V by summarizing our final results for the snake instability and offering a brief outlook toward future quantum mechanical calculations based upon our results. Appendix A will provide the detailed derivation of our results in Section IV. Appendix B will supply a pedagogical derivation of the analytical result for $k_{\text{max}}$ which is correctly stated without derivation in \cite{Kuznetsov}, and Appendix C will display a comparison between previous analytical formulas for the growth rate and our own findings.  

\section{Perturbation of a Condensate around a Gray Soliton}
\subsection{The gray soliton as a Gross-Pitaevskii solution}
In physical units the GP mean field equation of motion for the single-particle wave function into which many bosons have condensed is the nonlinear Schr\"odinger equation
\begin{equation}\label{GP0}
i\hbar\frac{\partial \Psi}{\partial t}=-\frac{\hbar^2}{2M}\nabla^2\Psi+\frac{4\pi a\hbar^2}{M}\vert \Psi\vert^2\Psi -\mu\Psi\;,
\end{equation}
where $M$ is the boson mass, $a$ is the scattering length for collisions between bosons, and $\mu$ is a chemical potential which may be freely shifted $\mu\to\mu+\Delta\mu$ by setting $\Psi\to e^{i\Delta\mu t}\Psi$ and is therefore conveniently tuned to make stationary solutions exactly time-independent. In a gas which is effectively confined to two dimensions $a$ is dimensionless, typically being given as the three-dimensional scattering length divided by the confinement length scale in the third direction. This equation has been validated experimentally for many aspects of the behavior of real Bose-Einstein condensates at very low temperatures, e.g. \cite{GPE_PEC1,GPE_PEC2}. The cubic nonlinear term $\vert \Psi\vert^2\Psi$ in the GP equation provides a mean-field description of short-ranged repulsive interaction between condensate particles. 

For any given typical gas density scale $\rho_0$ the GP equation has natural units based on the so-called \emph{healing length} $\xi = 1/\sqrt{4\pi a \rho_0}$, which in experiments is typically on the order of a micrometer. That is, we can rescale all our variables into dimensionless form using
\begin{align}
\mu &\to \frac{\hbar^2}{M\xi^2}\tilde{\mu}\nonumber\\
t &\to \frac{M\xi^2}{\hbar}\tilde{t}\nonumber\\
\mathbf{r} &\to \xi \tilde{\mathbf{r}}\nonumber\\
\Psi & \to\sqrt{\rho_0}\tilde{\Psi}\;.
\end{align}
Dropping the $\tilde{\ }$ accents for the remainder of this paper, the dimensionless GP equation then reads
\begin{equation}\label{GP}
i\frac{\partial \Psi}{\partial t}=-\frac{1}{2}\nabla^2\Psi+\vert \Psi\vert^2\Psi -\mu\Psi\;,
\end{equation}
where $\mu$ (now dimensionless) is still a tunable constant. 

A gray soliton is any member of the one-parameter family of one-dimensional macroscopic wavefunctions 
\begin{equation}\label{GS}
\Psi_{\beta}(x)=[\kappa\tanh(\kappa x)-i\beta]\,e^{i\beta x}
\end{equation}
which for all $-1\leq\beta\leq 1$ and $\kappa=\sqrt{1-\beta^{2}}$ are time-independent solutions to (\ref{GP}) if we set $\mu = 1+\beta^{2}/2$. The gray soliton represents a sort of nonlinear standing wave, in the form of an isolated `dip' in the condensate density:
\begin{equation}\label{GSden}
|\Psi_{\beta}|^{2}= (1-\beta^{2})\tanh^{2}(\kappa x) + \beta^{2} \equiv 1- \frac{\kappa^{2}}{\cosh^{2}(\kappa x)}\;.
\end{equation}
The term `gray' comes from nonlinear optical realizations of these solitons, in which the `dip' represents a small region of decreased light intensity. The special case $\beta=0$, where the minimum density reaches zero, is known as a `dark' soliton. For $\beta\not=0$, there is a steady non-zero flow of condensate through the density dip; the increased velocity needed to maintain the uniform flux through the low-density region means that the condensate phase $\theta(x)=\arg(\Psi_{\beta})$ makes a net jump $\Delta\theta = 2\cos^{-1}(\beta)$ across the dip.

Although $\Psi_{\beta}$ is a function solely of $x$, it is also a solution to the two-dimensional GP equation which is simply translationally invariant in the $y$ direction, so that the soliton's density `dip' becomes a trough extending along the $y$ axis. In one dimension the gray solitons remain robust, keeping their characteristic shape as they move over time, but in two dimensions they are linearly unstable to perturbations that break this $y$-translation invariance \cite{Snake1,Snake2}. The density trough spontaneously develops a snake-like `wiggle', as seen in Fig.~\ref{fig:GPev}. 

\subsection{Linearized perturbations}
A dilute Bose gas is still a quantum many-body system even when it is strongly Bose-condensed, and its exact description is in terms of second-quantized creation and destruction operator fields. The mean-field equation (\ref{GP}) is only a zeroth order approximation; quantum corrections are obtained by quantizing the perturbations around a solution to (\ref{GP}) as a classical background. As long as interactions are weak and the perturbations are small, however, it is a good approximation to neglect nonlinear dynamics in the time evolution of the perturbations. Nonlinear effects can then be included using quantum perturbation theory.

In this linear regime, moreover, quantization in the Heisenberg picture simply means taking classical solutions to linear evolution and letting their coefficients become time-independent operators. It is therefore a useful basis for further quantum studies to solve the linearized classical problem, leaving the introduction of quantum operators for later work. Such is our goal in this paper. In particular we focus on the snake mode as a class of perturbative normal modes, around the multi-dimensionally extended gray soliton, of which the linear evolution is dynamically unstable: instead of oscillating harmonically, these modes grow in time exponentially.

\subsection{Linear stability}
The linear stability of time-independent GP solutions may be determined by adding small perturbations to the time-independent solution (the `background' field), and then evolving the perturbed wave function under (\ref{GP}) while discarding terms of higher than first order in the perturbation. Since in our case the gray soliton background solution is independent of $y$, and we will be considering only terms linear in the perturbation to it, we can without loss of generality assume that the perturbation depends on $y$ sinusoidally:
\begin{equation}\label{sinky}
\Psi(x,y,t) = \Psi_{\beta}(x) + \delta\Psi(x,t)\cos(ky-\delta)\;.
\end{equation}
The linearization then consists of expanding Eqn.~(\ref{GP}) in $\delta\Psi$ and keeping only terms up to first order. This means setting
\begin{eqnarray}\label{BdG0}
i\frac{\partial \delta\Psi}{\partial t}&=&-\frac{1}{2}\frac{\partial^{2}\delta\Psi}{\partial x^{2}}+\left[2\vert\Psi_\beta\vert^2-\mu+\frac{k^{2}}{2}\right]\delta\Psi+\Psi_\beta^2\delta\Psi^{*}\;.
\end{eqnarray}
The system of coupled equations for $\delta\Psi$ and $\delta\Psi^{*}$ is known in general as the Bogoliubov-de Gennes equations (BdG); because our $\Psi_{\beta}$ is $y$-independent, they happen to take the especially simple form of partial differential equations in $t$ and $x$ only, with $k^{2}$ as an arbitrary real, positive parameter, even though they describe perturbations in two spatial dimensions. 

Since the BdG equations do not depend explicitly on $t$ either, their solutions can without loss of generality be decomposed further into eigenmodes of definite frequency, with the traditional notation being $\delta\Psi(x,t) = e^{-i\omega t}u(x)+e^{i\omega^{*}t}v^{*}(x)$. In our case, however, we are interested in dynamically unstable modes, with imaginary frequency $\omega\to i\lambda$ for $\lambda$ real. It is in general possible for BdG eigenfrequencies $\omega$ to be generally complex, but our numerical calculations have confirmed that the only complex $\omega$ in the BdG spectrum of the two-dimensional gray soliton are purely imaginary. Such growing or shrinking behavior in $\delta\Psi$ is possible even when we go beyond the c-number mean field theory to consider quantum fluctuations, because the quantized BdG equations do not correspond to Schr\"odinger equations, but rather to the linearization of a Heisenberg equation of motion for a quantum field. 

\subsection{The snake instability}\label{snake_inst}
For our case of dynamical instabilities with real growth rate $\lambda$, and translational symmetry in the $y$ direction, the common BdG notation with $u$ and $v^{*}$ is less convenient than writing
\begin{equation}\label{lininst}
\delta\Psi(x,t) = A e^{\lambda(k) t}e^{i\beta x}\phi_k(x)
\end{equation}
for some constant initial amplitude $A$, leaving one-dimensional time-independent BdG equations for $\phi_k(x)$:
\begin{eqnarray}\label{BdG1}
i\lambda\phi_k&=&-\frac{1}{2}\phi_k''-i\beta\phi_k'+\left(2\vert\Psi_\beta\vert^2-\mu+\frac{k^{2}+\beta^{2}}{2}\right)\phi_k\nonumber\\
&&\qquad+\Psi_\beta^2e^{-2i\beta x}\phi_k^{*}\nonumber\\
&=&-\frac{1}{2}\phi_k''-i\beta\phi_k'+\left(1-2\kappa^{2}\mathrm{sech}^{2}(\kappa x)+\frac{k^{2}}{2}\right)\phi_k\nonumber\\
&&\qquad +\left[\kappa\tanh(\kappa x)-i\beta\right]^{2}\phi_k^{*}\;.
\end{eqnarray}
Note the exact symmetry $\lambda\to-\lambda$, $\phi_k(x)\to\phi_k^{*}(-x)$. Both signs of $\lambda$ are always possible, as distinct imaginary eigenfrequencies, because our system is still Hermitian and has time reversal symmetry even though it is unstable, and so for every unstably growing mode there exists a time-reversed shrinking mode. Since the mode with negative $\lambda$ is obtained from the one with positive $\lambda$ just by reflection and complex conjugation, we will henceforth assume $\lambda(k)>0$.

The approximate analytical solution of (\ref{BdG1}) for small $k$ will be the subject of Section \ref{AnalyticalSection}, below. 
We can already anticipate the qualitative behavior of $\delta\Psi$, however, and also see why it is tempting to see the snake instability as a parametric one, just by looking at Fig.~\ref{fig:GPev}. It appears from Fig.~\ref{fig:GPev} that in the early stages of the instability we may have something much like
\begin{equation}\label{8eq}
\Psi(x,t) \sim \Psi_{\beta}\Big(x-A(t)\cos(ky-\delta)\Big)
\end{equation}
for some $A(t)$ that is initially small, because this represents a time-dependent modulation in $y$ of the $x$-coordinate of the density minimum---\textit{i.e.} the `snake'. By Taylor expanding (\ref{8eq}) in $A$, given the form of $\Psi_\beta$ from (\ref{GS}), we may then suspect a term of the form
\begin{equation}\label{9eq}
\phi_k(x) \sim \kappa^{2}\mathrm{sech}^{2}(\kappa x) \;
\end{equation} 
to appear in $\phi_k(x)$. As the amplitude $A(t)\sim A(0)e^{\lambda t}$ of the perturbation grows, the snake deformation increases. We will find that this simple expectation turns out to be correct up to, but only up to, a point.

\section{Analytical Theory of the Snake Instability}\label{AnalyticalSection}
\subsection{Finding $\phi_k(x)$ to first order in $k$}
In this Section we will derive $\phi_k(x)$ to leading order in small $k$, so that we obtain $\delta\Psi(x,y,t)$ according to (\ref{lininst}) in the limit of long wavelength `snaking'.  In our dimensionless units, this means that the approximation is accurate when the wavelength of the snake deformation in the $y$ direction is much longer than the ambient healing length of the condensate. Since for $k<0$ we can simply re-define $\delta\to\pi-\delta$ in (\ref{sinky}) to make $k\to-k$, we will take $k>0$ without loss of generality.

For $k=0$ our two-dimensional BdG problem reduces to the one-dimensional problem, which has been solved exactly for an arbitrary gray soliton \cite{Philip}. Since it has been shown in \cite{Philip} that there are no complex eigenfrequencies in the entire complete set of one-dimensional BdG excitations around any gray soliton, we know that if we expand $\lambda(k)$ in powers of dimensionless $k/\kappa$,
\begin{equation}\label{lambdak}
\lambda(k)=\kappa^2\sum_{n=0}^{\infty}\lambda_n\left(\frac{k}{\kappa}\right)^{n}\;,
\end{equation}
then we must have $\lambda_0=0$. Our main goals in this Section will be to compute $\lambda_1$ and to find the corresponding $\phi_k(x)$ to first order in $k/\kappa$, by using perturbation theory in $k/\kappa$ to extend the exact results for $k=0$ that are available in \cite{Philip}. This Section is thus a pedagogical review showing in detail how to obtain the results that were reported more briefly in Ref.~\cite{Kuznetsov}. In the following Section IV we will then extend our computation to further orders in $k/\kappa$ and obtain $\lambda_2$ and $\lambda_3$ as well as $\phi_k(x)$ up to second order in $k/\kappa$. These additional findings will improve the results, allowing analytical understanding of shorter-wavelength snake instabilities that are more rapid, but our higher-order corrections will also demonstrate the complexity of the `snaking' process at shorter snaking wavelengths.      

\subsection{The method of matched asymptotics}
We will need to use a somewhat more sophisticated perturbation theory than the basic kind, because the smallness of $k$ can manifest in two ways. It can make some terms in $\phi_k(x)$ small, such that they may be neglected; but it can also make some terms in $\phi_k(x)$ depend slowly on $x$, for example by depending on $kx$. Over a large enough range of $x$, the variation in $kx$ does not have to be small. If we merely perform the usual perturbative expansion in $k/\kappa$, therefore, writing something like 
\begin{equation}\label{phi_k}
\phi_k(x)=\sum_n^{\infty}\varphi_n(x)\left(\frac{k}{\kappa}\right)^n\;,
\end{equation}
our perturbation series may have a finite radius of convergence in $x$. One might think that this could be acceptable because we are mainly interested in what happens near the initial soliton, but the region near the soliton can in fact be affected significantly by deformations and waves that extend far away from it. A perturbative approach that only converges near the soliton will therefore fail to provide an accurate picture of what really happens even within this near region.

Regions far away from the soliton may thus pose a problem for the naive perturbation theory (\ref{phi_k}), but the problem is not that the distant regions are in themselves difficult. Far away from the soliton, our $\Psi_{\beta}(x)$ represents a uniform condensate with density 1 (in dimensionless units) and flow velocity $\beta$. The exact BdG solutions for small perturbations around such a background are well known for all values of the $y$-direction wave number $k$ and for any frequency $\omega$ including imaginary frequencies $i\lambda$. Far away from the soliton, therefore, we might not expect to need any perturbation theory in $k/\kappa$: we have exact BdG solutions asymptotically. Indeed the only difficulty in the large-$|x|$ `outer zone' is that the BdG equations are fourth order in spatial differentiation, so that our general solutions for $\phi_k(x\to-\infty)$ will have four undetermined coefficients, and those for $\phi_k(x\to+\infty)$ will have another four undetermined coefficients. These coefficients must be fixed by matching $\phi_k(x)$ smoothly from both sides into the `inner zone' of smaller $|x|$.

The inner and the outer zones thus present us with the problem that each of them influences the other. This problem is not a vicious circle, however, but rather an opportunity to construct a single global perturbation series in $k/\kappa$ for $\phi_k(x)$, converging everywhere. The reason we can do this is that the inner and outer zones \textit{overlap}. The `far away' outer zone in which we have asymptotically exact BdG solutions for all $k$ actually consists of all $|x|\gtrsim \kappa^{-1}$---not in general a long distance at all. This is because $\Psi_{\beta}$ as given by (\ref{GS}) approaches constant values exponentially fast with $\kappa |x|$. And on the other hand the inner zone `near the soliton', within which the $k$-perturbation theory will converge, consists of all $|x|\lesssim \sqrt{\frac{\kappa}{k}}$, so that $k \vert x\vert/\kappa$ remains small there as long as $k$ is small. For all $k\ll \kappa$, therefore, there will exist two large overlap regions $\kappa^{-1}\ll |x| \ll \sqrt{\frac{\kappa}{k}}$ (one region on each side of the soliton) within which both the perturbative and asymptotic limits apply. Within these overlap regions, then, we can compare the perturbative solution in the inner zone with the exact but under-determined outer zone solutions. By tuning the free parameters in each of these kinds of solutions in order to make them agree with each other, we can obtain a single global solution that is accurate everywhere, order by order in $k/\kappa$. One of the free parameters that will be tuned by this matching will be the growth rate $\lambda(k)$, which is thus also obtained as a series in $k/\kappa$. This procedure is an example of the \textit{method of matched asymptotics} \cite{MatchAsym}. 

We therefore look now first at the outer zone solutions, then at the inner solutions as given perturbatively by (\ref{phi_k}), and finally compare them and tune to find our global solutions.

\subsection{Outer zones}\label{section_oz}
In the zones far away from the soliton, we note that
\begin{equation}
\tanh(\kappa x) \equiv \mathrm{sgn}(x)\frac{1-e^{-2\kappa |x|}}{1+e^{-2\kappa |x|}}
\end{equation}
so that for the positive outer zone $x\gg\kappa^{-1}$ and negative outer zone $x\ll-\kappa^{-1}$, respectively, we can replace $\tanh(\kappa x)\to \pm1$ with exponentially small error. Neglecting these tiny errors, our time-independent BdG equation (\ref{BdG1}) becomes
\begin{eqnarray}\label{Outer0}
i\lambda\phi_k&=&-\frac{1}{2}\phi_k''-i\beta\phi_k'+\left(1+\frac{k^{2}}{2}\right)\phi_k\nonumber\\
&&\qquad +\left[\pm\kappa-i\beta\right]^{2}\phi_k^{*}\;,
\end{eqnarray}
where $\pm=\mathrm{sgn}(x)$. With the corresponding conjugate equation for $\phi_k^{*}$, this is a set of two coupled second-order ordinary differential equations, and within each positive and negative outer zone we can find the complete set of four linearly independent solutions by taking the Ansatz
\begin{equation*}
\phi_k(x)\, \overset{\scriptscriptstyle{x \to \pm\infty}}{\longrightarrow}\, (X_{\pm}+iY_{\pm})(\pm\kappa-i\beta)e^{-\gamma_{\pm}|x|}\;
\end{equation*} 
for real coefficients $X_{\pm}$ and $Y_{\pm}$, and $\gamma_{\pm}(k)$ that may in general be complex but will all turn out to be real for the small-$k$ cases we consider.
Inserting this Ansatz into (\ref{Outer0}) yields
\begin{eqnarray}
(\lambda\mp\beta \gamma_{\pm})X_{\pm}&=&\frac{k^2-\gamma_{\pm}^{2}}{2}Y_{\pm}\label{XY}\\
(\lambda\mp\beta \gamma_{\pm})Y_{\pm}&=&-\left(2+\frac{k^2-\gamma_{\pm}^{2}}{2}\right)X_{\pm}\label{YX}\;.
\end{eqnarray}
Multiplying these two equations by each other produces a quartic equation for $\gamma_{\pm}$:
\begin{equation}\label{quarticgamma}
\gamma_\pm^4 - (4\kappa^2+2k^2)\gamma_\pm^2 \mp 8\beta\lambda\gamma_\pm + (4k^2+k^4+4\lambda^2) = 0\;.
\end{equation}
 For each sign of $\pm$ however (\textit{i.e.}, for each of the two outer zones), two of the four roots of (\ref{quarticgamma}) have negative real parts, and must thus be discarded, because they imply un-normalizable $\phi_k(x)$ that diverge at infinity. 

The remaining two possible roots of (\ref{quarticgamma}) take simple limits whenever $k$ and therefore $\lambda(k)$ are small: one of them is $\gamma_{\pm}\doteq 2\kappa$ while the other is $\gamma_\pm$ of the same order as $k$ and $\lambda$. We can ignore the root $\gamma_{\pm}\doteq 2\kappa$, and keep only the smaller root for (\ref{quarticgamma}), because any terms in $\phi_k(x)$ which decay as $e^{-2\kappa |x|}$ will be entirely negligible in the outer zones $|x|\gg\kappa^{-1}$. 

We are thus left for $k\ll\kappa$ with only one possible spatial decay rate $\gamma_\pm$ on each of the $\pm$ sides of the soliton, and it is only unknown as a function of $k$ insofar as $\lambda(k)$ is unknown. If we apply (\ref{lambdak}) and solve (\ref{quarticgamma}) perturbatively, however, we can establish the small-$k$ limit
\begin{equation}\label{gamma}
\gamma_{\pm}=\left(\sqrt{\lambda_1^{2}+1}\mp \lambda_1\beta\right)\frac{k}{\kappa}+\mathcal{O}\left(\frac{k^{2}}{\kappa^2}\right)\;.
\end{equation}
Applying (\ref{YX}) to express $X_\pm$ in terms of $Y_\pm$, this leaves us with the outer zone solutions
\begin{eqnarray}\label{Outer1}
\kappa |x|\gg1\!:\quad&&\phi_k(x)=Y_{\pm}(\pm\kappa-i\beta)e^{-\gamma_{\pm}|x|}\\
&&\times \left(i-\frac{\lambda_1\mp\beta\sqrt{\lambda_1^{2}+1}}{2}\left(\frac{k}{\kappa}\right)+\mathcal{O}\left(\frac{k^{2}}{\kappa^{2}}\right)\right)\;\nonumber
\end{eqnarray}
which in the overlap region $\kappa^{-1}\ll |x|\ll \sqrt{\frac{\kappa}{k}}$ can be further expanded to
\begin{widetext}
\begin{align}\label{Outer2}
\kappa^{-1}\ll |x|\ll \sqrt{\frac{\kappa}{k}}\!:
&\quad\phi_k(x)=(\pm\kappa-i\beta)\left(\sum_{n=0}^{\infty}Y_{n\pm}\left(\frac{k}{\kappa}\right)^{n}\right)\left(1+\left(\lambda_1\beta\mp\sqrt{\lambda_1^{2}+1}\right)\left(\frac{k}{\kappa}\right)x+\mathcal{O}\left(\frac{k^{2}}{\kappa^2}\right)\right)\nonumber\\ &\quad\quad\quad\times\left(i-\frac{\lambda_1\mp\beta\sqrt{\lambda_1^{2}+1}}{2}\left(\frac{k}{\kappa}\right)+\mathcal{O}\left(\frac{k^{2}}{\kappa^{2}}\right)\right)\;
\end{align}
\end{widetext}
if we also expand the coefficient $Y_{\pm}$ in powers of $k/\kappa$. We will be able to fix $\lambda_1$, and also determine the full form of $\phi_k(x)$ to first order in $k/\kappa$, by comparing (\ref{Outer2}) with the corresponding result from the inner zone $|x|\ll \sqrt{\frac{\kappa}{k}}$, working order-by-order in $k/\kappa$.

\subsection{Inner zone and matching}\label{section_matching}
In the outer zone we made use of the restriction to small $k$ and $\lambda$, but our final retained $\gamma_{\pm}$ was kept as the exact root of the quartic that solved (\ref{Outer0}), and we only actually expanded perturbatively in powers of $k$ within the overlap zone $k|x|\ll 1$. In the inner zone, the more complicated form of the soliton background $\Psi_{\beta}$ only allows us to solve the BdG equation (\ref{BdG1}) exactly for $k=0$. The condition $k|x|\ll 1$ is true everywhere in the inner zone, however, and so we can apply ordinary perturbation theory for small but non-zero $k$. Formally we simply insert the $k$-expansion of $\phi_k(x)$ (\ref{phi_k}) into (\ref{BdG1}) with the corresponding expansion (\ref{lambdak}) for $\lambda(k)$, and solve for $\varphi_{n}(x)$ order-by-order in $k/\kappa$.

Ordinary perturbation theory is still somewhat more complicated for the time-independent BdG equations than for the single-particle Schr\"odinger equation. As a pair of coupled second-order differential equations for $\phi_k$ and $\phi_k^{*}$, the BdG system is effectively of fourth order. For the special case of BdG in the gray soliton background, however, the supersymmetric mapping introduced in \cite{Philip} allows an exact reduction of the BdG problem to second-order equations whose Green's functions can be obtained explicitly. This enables a straightforwardly algorithmic derivation of all the $\varphi_{n}(x)$ to arbitrary order $n$. Because this analysis is somewhat involved, we present it in detail in the Appendix A. In this Section we will simply present solutions, which can readily be checked by differentiation, as if they were obvious from inspection. 
At zeroth order in $k$, Eqn.~(\ref{BdG1}) reads
\begin{eqnarray}\label{inner0}
0&=&-\frac{1}{2}\varphi_{0}''-i\beta\varphi_{0}'+\left(1-2\kappa^{2}\mathrm{sech}^{2}(\kappa x)\right)\varphi_{0}\nonumber\\
&&\qquad +\left[\kappa\tanh(\kappa x)-i\beta\right]^{2}\varphi_{0}^{*}\;.
\end{eqnarray}
This has the general solution
\begin{eqnarray}
\varphi_{0}(x)&=&A_{0} \,\mathrm{sech}^{2}(\kappa x) + \frac{B_{0}}{\kappa}[i\kappa\tanh(\kappa x)+\beta]\\
&&+C_{0}\Big(i x[\kappa\tanh(\kappa x)-i\beta] -i\nonumber\\
&&\qquad\qquad-\frac{3\beta}{2\kappa}[\kappa x\,\mathrm{sech}^{2}(\kappa x) +\tanh(\kappa x)]\Big)\nonumber
\end{eqnarray}
for real $A_{0}$, $B_{0}$, $C_{0}$, after discarding a fourth solution which grows as $e^{+2\kappa|x|}$ and will therefore never be able to match smoothly onto our solution from the outer zone. In the overlap regions $\pm\kappa x\gg1$ this $\varphi_{0}(x)$ becomes
\begin{align}
&\pm\!\kappa x \gg1\!:\\
&\quad\varphi_{0}(x)\to \frac{B_{0}}{\kappa}(\pm i\kappa+\beta)+C_{0}\Big[(\pm i\kappa+\beta) x - i \mp \frac{3\beta}{2\kappa}\Big]
\;.\nonumber
\end{align}
We now compare this with the $\mathcal{O}\left(\left(k/\kappa\right)^{0}\right)$ term in the expansion (\ref{Outer2}) of our outer zone solution in the overlap region, which was simply $(\pm i\kappa +\beta)Y_{0\pm}$. From the absence of any terms $\sim x$ in the outer zones at zeroth order, we conclude that we must have $C_{0}=0$. Matching the remaining terms fixes $Y_{0+}=Y_{0-}=B_{0}/\kappa$. The inner zone coefficient $A_{0}$ remains undetermined; in fact it will remain arbitrary, since the BdG equations are linear equations that admit an arbitrary overall constant prefactor in their solutions. All the other coefficients will be fixed in relation to $A_{0}$, when we pursue the matching at higher orders in $k/\kappa$.

At first order in $k/\kappa$, Eqn.~(\ref{BdG1}) says that
\begin{eqnarray}\label{inner1}
i\kappa^2\lambda_1\varphi_{0}&=&-\frac{1}{2}\varphi_{1}''-i\beta\varphi_{1}'+\left(1-2\kappa^{2}\mathrm{sech}^{2}(\kappa x)\right)\varphi_{1}\nonumber\\
&&\qquad +\left[\kappa\tanh(\kappa x)-i\beta\right]^{2}\varphi_{1}^{*}\;.
\end{eqnarray}
This has the solution
\begin{eqnarray}
\varphi_{1}(x)&=&-\lambda_1\left(\frac{2\beta A_{0}+B_{0}\kappa}{2\kappa}\right)\Big(\kappa x \,\mathrm{sech}^{2}(\kappa x)+\tanh(\kappa x)\Big)\nonumber\\
&&-i\lambda_1A_{0}\nonumber\\
&&+A_{1} \,\mathrm{sech}^{2}(\kappa x) + \frac{B_{1}}{\kappa}[i\kappa\tanh(\kappa x)+\beta]\\
&&+C_{1}\Big(i x[\kappa\tanh(\kappa x)-i\beta] -i\nonumber\\
&&\quad\qquad-\frac{3\beta}{2\kappa}[\kappa x\,\mathrm{sech}^{2}(\kappa x) +\tanh(\kappa x)]\Big)\nonumber \;.
\end{eqnarray}
which in the overlap regions becomes
\begin{align}\label{match1A}
&\pm\kappa x \gg1\!:\nonumber\\
&\quad\varphi_{1}(x)\to \frac{1}{\kappa}(B_{1}\mp\lambda_1 A_{0})(\pm i\kappa+\beta)\mp\frac{\lambda_1}{2}B_{0}\nonumber\\
&\qquad+C_{1}\Big[(\pm i\kappa+\beta) x - i\mp \frac{3\beta}{2\kappa}\Big]\;.
\end{align}

The $\mathcal{O}(k/\kappa)$ term in (\ref{Outer2}) implies, however, that in the overlap regions we must have
\begin{eqnarray}\label{match1B}
\varphi_{1}(x) &=& (\pm i\kappa +\beta)\Bigg(Y_{1\pm} +i\frac{B_{0}}{\kappa}\frac{\lambda_1\mp\beta\sqrt{\lambda_1^{2}+1}}{2}\nonumber\\
&&\qquad\qquad+\frac{B_{0}}{\kappa}\left(\lambda_1\beta\mp\sqrt{\lambda_1^{2}+1}\right)x\Bigg)\;.
\end{eqnarray}
Matching in the two distinct overlap regions means that Eqns.~(\ref{match1A}) and (\ref{match1B}) must agree for both signs of $\pm$. Looking at the terms containing $x$ in each expression, we see that the two $\pm$ cases impose incompatible conditions on $C_{1}$ and $B_{0}$, which must therefore both vanish. The remaining terms then agree if and only if we set $Y_{1\pm}=(B_{1}\mp \lambda_1A_{0})/\kappa$.

\subsection{Determining the growth rate $\lambda_1$}
The calculation of $\lambda$ is based on taking the real part of equation (\ref{BdG1}), multiplying it by $\mathrm{sech}^2(\kappa x)$, and integrating over all $x$. This yields \begin{widetext}
\begin{align}\label{RHSvanisch}
\lefteqn{\int\limits_{-\infty}^{\infty}\!\mathrm{dx}\, \mathrm{sech}^2(\kappa x)\left(\lambda\,\mathrm{Im}(\phi_k(x))+\frac{k^2}{2}\,\mathrm{Re}(\phi_k(x))\right)}&\nonumber\\
&= \int_{-\infty}^{\infty}\!\mathrm{dx}\,\mathrm{sech}^{2}(\kappa x)\left[\left(\frac{1}{2}\mathrm{\frac{d^{2}}{dx^{2}}}-\kappa^{2}[2-3\,\mathrm{sech}^{2}(\kappa x)]\right)\mathrm{Re}(\phi_k)-2\beta\left(\mathrm{\frac{d}{dx}}-2\kappa\tanh(\kappa x)\right)\mathrm{Im}(\phi_k) \right]\nonumber\\
&\equiv \int_{-\infty}^{\infty}\!\mathrm{dx}\,\mathrm{\frac{d}{dx}}\left[\frac{1}{2}\mathrm{sech}^{2}(\kappa x)\,\mathrm{\frac{d}{dx}}\mathrm{Re}(\phi_k) +\kappa\,\mathrm{sech}^{2}(\kappa x)\,\tanh(\kappa x)\mathrm{Re}(\phi_k)-2\beta\,\mathrm{sech}^{2}(\kappa x)\,\mathrm{Im}(\phi_k) \right]\equiv 0\;.
\end{align}
\end{widetext}

Note that our expressions from matched asymptotics for the functions $\phi_k(x)$ will smoothly combine inner and outer zone solutions, and hence be valid over the full infinite range of $x$ integration in (\ref{RHSvanisch}). Because $\mathrm{sech}^{2}(\kappa x)$ decays exponentially for large argument, however, this factor in the integrand on the left-hand side ensures that only the inner zone part of the solution for $\phi_k$ is needed to determine $\lambda(k)$. Inserting our power series in $k/\kappa$ (\ref{lambdak}) and (\ref{phi_k}) for both $\lambda(k)$ and $\phi_k(x)$ into (\ref{RHSvanisch}), we obtain a recursion relation for $\lambda_n$ involving $\lambda_{m\leq n-1}$ and integrals of $\mathrm{Im}(\varphi_{m\leq n})$ and $\mathrm{Re}(\varphi_{m\leq n-1})$. 

For $\lambda_1$ we can insert our matched results so far for $\varphi_{0}(x)$ and $\varphi_{1}(x)$ into (\ref{RHSvanisch}) 
\begin{align}\label{order2a}
\int_{-\infty}^{\infty} \mathrm{sech}^2(\kappa x)\left(2\lambda_1\mathrm{Im}(\varphi_{1})+\mathrm{Re}(\varphi_{0})\right)\mathrm{d}x=0
\end{align}
and find
\begin{widetext}
\begin{align}
0=&\int_{-\infty}^{\infty}\left(2\lambda_1\left(\lambda_1A_0-B_1\tanh(\kappa x)\right)\right)\mathrm{sech}^2(\kappa x)\mathrm{d}x
-\int_{-\infty}^{\infty}A_0\,\mathrm{sech}^4(\kappa x)\mathrm{d}x\;.
\end{align} 
\end{widetext}
The still unknown constant $B_1$ does not pose a problem here, since the integral containing $B_1$ vanishes. Performing the remaining integral determines $\lambda_1$ as found by essentially this same method in Ref.~\cite{Kuznetsov}:
\begin{align}\label{lambda1}
\lambda_1=\frac{1}{\sqrt{3}}\;.
\end{align}
As we noted in Subsection \ref{snake_inst} above, both signs of $\lambda$ are always possible because our system is Hermitian. We consider only the positive branch of $\lambda_1$, however, because the shrinking mode with negative $\lambda$ can always be obtained by taking $\phi_k(x)\to\phi_k^{*}(-x)$.

\subsection{Higher order matching}
To achieve our further goal of determining not only $\lambda$ but also $\phi_k(x)$ to first order in $k/\kappa$, we actually need to consider the next higher order of expansion of (\ref{BdG1}) in $k/\kappa$, namely the quadratic order. At second order in $k/\kappa$ Eqn.~(\ref{BdG1}) reads
\begin{widetext}
\begin{eqnarray}\label{order2}
	\kappa^2\left(i\lambda_2-\frac{1}{2}\right)\varphi_{0}(x)+i\kappa^2\lambda_1\varphi_{1}(x) &=&-\frac{1}{2}\varphi_{2}''-i\beta\varphi_{2}'+\left(1-2\kappa^{2}\mathrm{sech}^{2}(\kappa x)\right)\varphi_{2}+\left[\kappa\tanh(\kappa x)-i\beta\right]^{2}\varphi_{2}^{*}\;.
\end{eqnarray}
\end{widetext}

The as-yet-undetermined constant $B_{1}$ in $\varphi_{1}(x)$ can then be fixed by solving (\ref{order2}) for $\varphi_{2}(x)$ and again matching with the $\mathcal{O}(k^{2}/\kappa^2)$ term in the outer zone solution (\ref{Outer2}). This matching is made simpler by the fact that we have already determined $Y_{0\pm}=B_{0}=0$. It is typical of the method of matched asymptotics to find that some coefficients at one order are only fixed by matching at the next order \cite{MatchAsym}.

The only remaining unknown coefficients in $\varphi_{1}$ and $\varphi_{0}$ are $A_{1}$ and $A_{0}$. Since both multiply the same $\mathrm{sech}^{2}(\kappa x)$ term, and since an overall rescaling $\phi_k(x)\to \tilde{A}\phi_k(x)$ is always allowed by the linear BdG equations, we can consider $\tilde{A}=1-kA_{0}^{-1}A_{1}$ and so set $A_{1}$ to zero without loss of generality. $A_{0}$ then remains free as the overall excitation amplitude of the unstable snake mode with transverse wave number $k$. 

\subsection{Global solution to order $k$}\label{global_k}
Now that we have found smoothly matching solutions in the inner and outer zones, we can put them together into a global solution for all $x$ up to first order in $k/\kappa$. The usual procedure when using matched asymptotics approach would be to put a border cut at any point within the overlap zone and then construct a piecewise-defined function, consisting of the two previously found solutions. The resulting function then represents the global solution. 

In our particular case, however, this simple `patching' approach will turn out in the next Section to be unsatisfactory at higher orders in $k$, because our inner-zone solution $\varphi_2(x)$ will turn out to include terms which grow linearly in $x$ at large $x$. These terms must be present in order to agree, in the overlap region, with the exponentially decaying terms $e^{-\gamma_{\pm}\vert x\vert}$ of the outer zone: matching order-by-order in $k$ implies Taylor expanding the outer-zone exponential within the overlap region, generating the linearly growing terms in the matched inner zone solution. This unfortunately means, however, that order-by-order matching of the inner and outer solutions does not really yield smooth matching, because a linear function of $x$ is not really like an exponential function of $|x|$ even when the Taylor expansions of the two functions are matched at low order. As a result, the patched-together function will always have sharp corners or discontinuities that are formally of higher order in $k$ but that are qualitatively wrong, since the actual global solution is smooth.

Although this problem will not actually arise until the higher-order correction $\varphi_{2}$ that we will compute in the next Section, we introduce here the modified patching procedure that we will use to obtain a smooth global solution from our matched asymptotic results. The procedure is to define the global solution $\phi_k(x)$ as the product of an envelope that reproduces the outer zone solution and a function $\tilde{\phi}_{k}(x)$ which is chosen to ensure that the product function $\phi_k(x)$ correctly reproduces our inner zone solution for $x$ in the inner zone, but also itself becomes \emph{constant} at large $|x|$ rather than containing any secularly growing terms. That is:
\begin{align}\label{trafo}
\phi_k(x)&=\tilde{\phi}_k(x)e^{\frac{k}{\kappa}h(x)}\qquad\hbox{for}\\
\frac{k}{\kappa}h(x)&=-\frac{\gamma_+ -\gamma_-}{2}x-\frac{\gamma_+ +\gamma_-}{2}x\tanh(\kappa x)\nonumber\\
&\doteq\left(\lambda_1\beta x-\sqrt{\lambda_1^2+1}\;x\tanh(\kappa x)\right)\frac{k}{\kappa}\nonumber\\
&=\frac{kx}{\sqrt{3}\kappa}[\beta - 2 \tanh(\kappa x)]+\mathcal{O}(k^2)\;.
\end{align}   
This envelope construction explicitly ensures the correct asymptotic behavior in the outer zone, since $\lim_{\vert x\vert \to \infty} e^{h(x)\frac{k}{\kappa}} = e^{-\gamma_{\pm}\vert x\vert}$. 

Choosing $\tilde{\phi}_k(x)$ to give $\phi_k(x)$ the correct form in the inner zone after the envelope factor is included is a systematic procedure based simply on Taylor-expanding the envelope; in our next Section it will be continued to second order with
\begin{align}\label{trafo2}
\tilde{\phi}_k(x)=\;&\varphi_0+\Big(\varphi_1(x)-\varphi_0(x) h(x)\Big)\left(\frac{k}{\kappa}\right)\nonumber\\
&+\Big(\varphi_2(x) - \varphi_1(x)h(x) + \frac{1}{2}\varphi_0(x)h^2(x)\Big) \left(\frac{k}{\kappa}\right)^2\nonumber\\
& + \mathcal{O}\left(\frac{k}{\kappa}\right)^3\;.
\end{align}
This procedure also ensures that $\tilde{\phi}_k(x)$ becomes constant at large $|x|$, because the growing terms in the inner zone solution are in fact nothing but Taylor expansions of the outer-zone exponentials, by which the inner zone solution matches the outer zone solution, in the overlap region, order by order in $k/\kappa$.

(In principle we could simply have used (\ref{trafo}) to define~$\tilde{\phi}_k(x)$, then derived the differential equation satisfied by~$\tilde{\phi}_k(x)$ from Eqn.~(\ref{BdG1}) for $\phi_k(x)$, and finally solved for~$\tilde{\phi}_k(x)$ directly through a perturbative approach just like the matched asymptotics procedure we use to obtain~$\varphi_n(x)$, except with the boundary condition of asymptotic constancy at large $|x|$ instead of matching with the outer zone. This equivalent alternative procedure may be conceptually simpler than deriving the~$\varphi_n(x)$ by matched asymptotics, as we have actually done, and then adjusting the envelope via (\ref{trafo}) and (\ref{trafo2}). However, this conceptually simpler approach turns out to be considerably more complicated in execution than our less elegant approach with (\ref{trafo}) and (\ref{trafo2}), because the differential equation satisfied by $\tilde{\phi}_k(x)$ is much more tedious to solve than Eqn.~(\ref{BdG1}). In practice it proves to be easier to first obtain a working solution by the cruder method of matched asymptotics and patching, and then refine the solution's appearance by the procedure (\ref{trafo2}) that is straightforward once the solution is known.)\\

\subsection{A parametric instability?}
\subsubsection{First main result}
Our final step at order $k/\kappa$, namely finding $B_{1}$, has already gone beyond the solution for (what in our notation is) $\varphi_{1}$ that was offered in Ref.~\cite{Kuznetsov}, where the $B_{1}$ term was ignored because it played no role in determining $\lambda_1$. It is interesting to note, though, that the $\delta\Psi$ which is now fully given to first order in $k/\kappa$ by our $\varphi_{1}$ is still composed of terms which are proportional to the unperturbed gray soliton wave function $\Psi_{\beta}(x)$, or to its derivatives with respect to $x$ or $\beta$. This implies that, at least to linear order in $k/\kappa$, the snake mode instability is indeed a parametric instability, for we can write\begin{widetext}
\begin{align}
\Psi(x,y,t) &= \Psi_{\beta}(x) + \delta\Psi(x,y,t)=\Psi_\beta(x) + e^{\lambda t}e^{i\beta x}\cos(ky-\delta)\left[\varphi_0(x)+\frac{k}{\kappa}\varphi_1(x) + \mathcal{O}\left(\frac{k}{\kappa}\right)^2\right]\label{nonparam}\\
&=e^{i\beta x}e^{i\theta_C(x,y,t)}e^{-iS(x,y,t)}\left(\tilde{\kappa}\tanh\big(\tilde{\kappa}(x-Q)\big) -i\tilde{\beta}\right)+\mathcal{O}\left(\frac{k}{\kappa}\right)^2+\mathcal{O}(A_0^2)\;,\label{param1}
\end{align}\end{widetext}
for
\begin{align}
Q(y,t) &= A_0 e^{\lambda t}\cos(ky-\delta)\\
\tilde{\beta}(y,t) & = \beta + \frac{k}{\sqrt{3}\kappa} Q(y,t)\nonumber\\
\tilde{\kappa}(y,t) &= \sqrt{1-\tilde{\beta}^2}\nonumber\\
\theta_C(x,y,t) & =Q(y,t)\frac{k}{\sqrt{3}\kappa^2}\tanh(\kappa x)\nonumber\\
S(x,y,t) & = Q(y,t)\frac{k}{\sqrt{3}\kappa^2}[\tanh(\kappa x)+\beta]e^{\frac{k}{\kappa}h(x)}\;.\nonumber
\end{align}
That is, if we expand the expression in (\ref{param1}) up to first order in $A_0$, it agrees exactly with our matched asymptotics results in (\ref{nonparam}), to first order in $k/\kappa$.

The functional form of $\Psi_\beta+\delta\Psi$ that we have found up to linear order in $k/\kappa$ is thus still locally the form of a gray soliton, but the position and $\beta$ parameter of the soliton are shifted in ways that depend linearly on $A_0$, and sinusoidally on $y$ and exponentially on $t$. The $y$-dependent displacement and deformation of the soliton are also accompanied by a sound wave represented by the phase perturbation $S(x,y,t)$. Since this long-range sonic `dressing' of the snake mode decays exponentially in $x$ in both directions away from the soliton at the same time that it grows exponentially in time, it propagates away from the soliton at the speed of sound (which is anisotropic in the lab frame because of the background phase flow $e^{i\beta x}$). 

\subsubsection{Phase counterterm}
It should be noted that there is \emph{no} long-range exponential decay factor $\exp(kh(x)/\kappa)$ in the separate phase term $\theta_C(x,y,t)$ which appears in (\ref{param1}). This phase modulation $\theta_C$ is simply part of our $\varphi_1(x)$ solution, and not something extra which has to be added by hand, but it lacks the long-range matching factor $\exp(kh(x)/\kappa)$, and is written separately from the rest of the phase dressing $S(x,y,t)$, because $e^{i\theta_C(x,y,t)}$ plays the role of an infra-red counterterm to the $\beta\to\tilde{\beta}(y,t)$ perturbation. Shifting the position of the soliton with $Q(y,t)$ has no effect on the order parameter $\Psi(x,y,t)$ at large $x$, but shifting $\beta$ changes the phase of $\Psi$ at long range, and it changes it in opposite directions for $x\to\pm\infty$. Since the shift $\beta\to\tilde{\beta}$ is also $y$-dependent, shifting $\beta$ in this way introduces a non-trivial phase wave that extends to infinity. Actually changing $\Psi$ in that way would represent instantaneous action at a distance in the snake mode, and in an infinite sample it would also cost infinite energy (because of the $y$-dependence). The $\theta_C(x,y,t)$ term does very little near the soliton for small $k$, but at large $|x|$ it exactly cancels the phase changes introduced by the perturbation to $\beta$, so that the combination of linearly perturbing $\beta\to\tilde\beta$ and adding $\theta_C$ to the phase has a net effect on $\Psi$ which is exponentially localized near the soliton. Thus the derivation of the asymptotic matching factor $\exp(kh(x)/\kappa)$ in Subsection G above, from considerations of convergence of the perturbation series in $k$ when $k/\kappa$ is small but $kx$ is not, does not affect $\theta_C$ because although $\theta_C(x,y,t)$ itself extends to infinity, it is actually just one part, together with the $\beta\to\tilde\beta$ deformation, of a perturbation which does not extend to large $kx$.

\subsubsection{Value of parametric form}
Recognizing that the snake mode is a parametric deformation (dressed by a sound wave) would be difficult to do just by inspecting numerical solutions; discoveries like this are an important benefit that can still be gained from analytical calculations. This discovery is useful because parametric instabilities can often be understood beyond the limit of small perturbations, by considering the shifting parameters as collective coordinates which can be allowed to vary by finite amounts, in what amounts to a well-motivated time-dependent variational Ansatz. Here, for example, the collective coordinate would be $Q(y,t)$; all other changes in $\Psi$ are proportional to it.

This approach can even be pursued quantum mechanically, by quantizing the collective coordinate within a much smaller Hilbert space than that of the full many-body system, delivering a tractable quantum theory that can be compared with experiments. It is thus important to ask whether the gray soliton snake instability is still a parametric instability at larger $k$, where $\lambda$ is larger and the faster-growing instabilities will therefore tend to dominate the longer-wavelength snake modes that are described by our solution to linear order in $k/\kappa$. We will therefore now continue further beyond Ref.~\cite{Kuznetsov} by going to higher orders in $k/\kappa$.

\section{Going beyond first order in $ k $}
\subsection{Results to order $k^2$}
In this Section we will extend our calculation of $\varphi_{n}(x)$ up to the third order in $k/\kappa$ using similar methods to those introduced in the previous Section. Our detailed calculations can be found in our Appendix A; in this Section we will simply present the solutions. 

Solving Eqn.~(\ref{order2}) and performing the matching with the outer zone yields (see Appendix) this result for $\varphi_{2}(x)$:
\begin{widetext}
\begin{align}\label{phi2A}
\varphi_2(x)=&-\left(\frac{2\beta A_0\lambda_2+B_1\kappa\lambda_1}{2\kappa}\right)\left(\kappa x \,\mathrm{sech}^2(\kappa x)+\tanh(\kappa x)\right)+\left(i\frac{B_2}{\kappa}-A_0 \lambda_1^2 x\right)\left(\kappa\tanh(\kappa x)-i\beta\right)\nonumber\\
&+\frac{A_0 \lambda_1^2}{\kappa}x\tanh(\kappa x)+\frac{1}{2}A_0 \lambda_1^2\beta^2 x^2\,\mathrm{sech}^2(\kappa x)+\frac{A_0}{6}-iA_0\lambda_2-\frac{A_0\lambda_1^2\beta^2}{2\kappa^2}\nonumber\\
&+A_2\,\mathrm{sech}^2(\kappa x)+C_2\,\Big(i x[\kappa\tanh(\kappa x)-i\beta] -i-\frac{3\beta}{2\kappa}[\kappa x\,\mathrm{sech}^{2}(\kappa x) +\tanh(\kappa x)]\Big)\;.
\end{align}\end{widetext}
We show in Appendix A that matching in both outer zones is a strong enough condition to establish 
\begin{align}\label{B1C2}
B_{1}&=-2 A_0\beta\lambda_1^2/\sqrt{\lambda_1^2+1} = -\frac{\beta}{\sqrt{3}}A_0\nonumber\\
C_{2}&=\frac{\lambda_1}{\kappa}\left(\beta B_1+\sqrt{\lambda_1^2+1}\,A_0\right)=\frac{2-\beta^2}{3\kappa}A_0\;.
\end{align}
The co-efficient $B_2$ can only be fixed by matching at order $(k/\kappa)^3$, just as $B_1$ was obtained at order $(k/\kappa)^2$, but we also show in Appendix A that it turns out to be
\begin{align}
B_2=-\frac{11}{24}A_0\beta\kappa\;.
\end{align}

Even before knowing $B_{2}$ we can determine $\lambda_2$ from (\ref{RHSvanisch}), since just as with $B_{1}$, the unknown $B_{2}$ does not contribute to the integrals in (\ref{RHSvanisch}). We therefore use
\begin{align}\label{order3a}
\int\!\mathrm{dx}\, \mathrm{sech}^2(\kappa x)\left(\lambda_1\mathrm{Im}(\varphi_{2})+\lambda_2\mathrm{Im}(\varphi_{1})+\frac{\mathrm{Re}(\varphi_{1})}{2}\right)=0
\end{align}
to obtain
\begin{align}\label{lambda2}
\lambda_2=-\frac{1+\beta^2}{6\kappa}=-\frac{2-\kappa^2}{6\kappa}\;.
\end{align}
This second-order result has been reported previously \cite{KP2000,KP2008} without explicit derivation; in Appendix C we discuss these earlier approximate treatments of $\lambda(k)$. We focus now on our main goal of studying how $\Psi(x,y,t)$ evolves as the snake instability grows.

\subsection{Still parametric ... almost}
Inserting our results for the coefficients, and tuning $A_2$ to cancel any $\mathrm{sech}^2$ term since it can still be absorbed into $A_0$, we can rearrange terms in (\ref{phi2A}) and extend our parametric expression (\ref{param1}) into
\begin{widetext}
\begin{align}
\Psi(x,y,t) &= \Psi_{\beta}(x) + \delta\Psi(x,y,t)=\Psi_\beta(x) + e^{\lambda t}e^{i\beta x}\cos(ky-\delta)\left[\varphi_0(x)+\left(\frac{k}{\kappa}\right)\varphi_1(x) + \left(\frac{k}{\kappa}\right)^2\varphi_2(x)+ \mathcal{O}\left(\frac{k}{\kappa}\right)^3\right]\nonumber\\
&=e^{i\beta x}e^{i\theta_C(x,y,t)}e^{-iS(x,y,t)}\tilde\rho(x,y,t)\left(\tilde{\kappa}\tanh\big(\tilde{\kappa}(x-Q)\big) -i\tilde{\beta}\right)+\Phi_\mathrm{NP}(x,y,t)+\mathcal{O}\left(\frac{k}{\kappa}\right)^3+\mathcal{O}(A_0^2)\;,\label{param2}
\end{align}
still for $Q(y,t) = A_0 e^{\lambda t}\cos(ky-\delta)$ and $\tilde{\kappa}=\sqrt{1-\tilde{\beta}^2}$ as before, but now with the extended and additional parameter values that are given up to corrections of order $(k/\kappa)^3$ by
\begin{align}\label{param3}
\tilde{\beta}(y,t) & = \beta + Q(y,t)\left[\frac{k}{\sqrt{3}\kappa}+\frac{(3-4\beta^2)k^2}{6\kappa^3}\right] \nonumber\\
\theta_C(x,y,t) & =Q(y,t)\left[\frac{k}{\sqrt{3}\kappa^2}+\frac{(3-4\beta^2)k^2}{6\kappa^4}\right] \tanh(\kappa x)\nonumber\\
S(x,y,t) & = Q(y,t)e^{\frac{k}{\kappa}h(x)}\left[\frac{k}{\sqrt{3}\kappa^2}[\tanh(\kappa x)+\beta]\right.\nonumber\\
&\qquad\qquad\left.+\frac{k^2}{6\kappa^4}\left((3-2\beta^2)\tanh(\kappa x)+\frac{(7-3\beta^2)\beta}{4}-[(\beta^2+1)\kappa x + 2\beta]\,\mathrm{sech}^2(\kappa x)\right)\right]\nonumber\\
\tilde{\rho}(x,y,t)&=Q(y,t)\left[1+\frac{k^2}{6\kappa^3}[\tanh(\kappa x)+\beta][1-2\beta\tanh(\kappa x)]\right]\nonumber\\
\Phi_{NP}(x,y,t)&=Q(y,t)\frac{k^2}{6\kappa^3}\,\mathrm{sech}^2(\kappa x)\left[\beta^2\kappa x^2 + (3-\beta^2)i \kappa x \tanh(\kappa x) + 2\beta(i+\kappa)[\kappa x - \tanh(\kappa x)]\right]
\;.
\end{align}\end{widetext}
Here our previous parametric deformations proportional to the single collective coordinate $Q(y,t)$ are all present, just with further terms at order $k^2/\kappa^2$. The soliton is shifted in position and grayness parameter $\beta$, and also dressed by an exponentially decaying sound wave. The new factor $\tilde{\rho}(x,y,t)$ is actually part of this sound-wave dressing as well: long-wavelength waves of BdG `zero sound' in a uniform background, to which our background $\Psi_\beta$ reduces away from the soliton, are waves of condensate phase and density, with the density modulation proportional to the phase wave but smaller by a factor of $k$. 

The final term $\Phi_{NP}(x,y,t)$, however, is a new correction which must be added at second order in snaking wave number $k$. It cannot be represented as a change of the background within the two-parameter family of gray solitons, modulated by a sound wave; instead it describes a change of the order parameter away from soliton form. We can note, however, that $\Phi_{NP}$ is not only of order $(k/\kappa)^2$, but moreover consists of terms that vanish for small $x$ as $x^2$ or even $x^3$, multiplied by the $\mathrm{sech}^2(\kappa x)$ term which vanishes exponentially for large $x$. $\Phi_{NP}$ is thus everywhere small. A parametric Ansatz for the snake mode which simply neglects $\Phi_{NP}$ will probably be sufficiently accurate for most purposes, as long as the snaking wave number $k$ is not too large.

It is therefore still fair to say that the snake instability remains essentially a parametric instability of the gray soliton up to order $k^2$, sustaining the hope that future quantum calculations may be able to be based on quantizing a collective coordinate. Unfortunately, however, this trend does not continue to all orders in $k$.

\subsection{Order $k^3$: a new functional form}
Although the snake mode may still be approximately parametric at order $k^2$, the small non-parametric correction $\Phi_{NP}$ does exist. The parametric deformation has also become rather complex in its dependences on $x$ and on $\beta$. Even without performing further calculations we might suspect at this point that the snake mode will not remain parametric to all orders in $k$. And indeed it does not.

In a similar way to our analysis so far, we can now go on to calculate $\varphi_{3}$ and $\lambda_3$. The full expression for $\varphi_{3}(x)$ can be found in Appendix A, with $B_{3}$ once again as an undetermined constant. In principle we could indeed keep going in this way indefinitely; but (as can be seen in the Appendix) terms with known constant coefficients in the third order solution $\varphi_{3}$ include dilogarithms, $\mathrm{Li}_2(-e^{-2 x\kappa})$. These less-well-known functions lie outside the familiar family of hyperbolic functions that were contained in $\varphi_{0,1,2}$. Their addition makes the analytical calculation of all subsequent orders of $\varphi_{n}(x)$ much more difficult, as even more exotic special functions will presumably accumulate at higher orders.

In spite of the presence of the dilogarithms at third order in $k$ we can still perform the matching with the outer zone to fix the coefficient $B_2$ that remained unknown at second order; we find the value shown in the previous subsection, and after thus fixing $B_2$ we have determined the second order of $\varphi_{n}(x)$ completely. In a similar way, however, our new third-order coefficient $B_3$ remains undetermined at order $k^3$. Fixing it to complete the determination of $\varphi_3(x)$  would require finding the solution for $\varphi_{4}(x)$ and again perform the matching with the outer zone at this order. The effort to do this, however, would outweigh the benefits won from this calculation. All we really need to know about $\varphi_3(x)$ itself is that the presence of the unfamiliar dilogarithm function makes it opaque enough that at this point one might as well just rely on numerical solutions. And $\varphi_4(x)$ can only be worse.

The fact that $\varphi_3(x)$ contains terms that do not look like parametric modulations of the background gray soliton is potentially discouraging news for future quantum calculations. The hope of describing finite-amplitude snake excitations within a variational subspace relies on being able to assume a sufficiently accurate Ansatz even for nonlinear deformations of the soliton; if the deformation of the soliton is becoming this complicated even within the linearized BdG theory, then it is unclear what kind of variational Ansatz should be preferred for deformations beyond the linear regime. Constructing heuristic theoretical models to compare with experiment may be a worthwhile approach to a problem as difficult as nonlinear quantum many-body evolution, but even quantized collective coordinate models are difficult enough to analyze that one does not wish to base them on a variational Ansatz guessed simply at random.

This suggests that future efforts at understanding the quantum snake mode might base themselves on $\Psi$ up to order $k^{2}$ alone, as given above in (\ref{param2}), constructing a variational Ansatz from the parametric deformations which this restricted $\Psi$ represents and quantizing within the reduced Hilbert space. How well would the results of such a `moderately long-wavelength' theory represent the most rapidly growing instabilities that would be expected to dominate in experiments? To estimate an answer to this question in advance, we will look in subsection \ref{errors} at how well the quadratic approximation $\lambda(k)=\kappa\lambda_1k + \lambda_2k^{2} + \mathcal{O}(k^{3})$ compares with numerically obtained curves for the full $\lambda(k)$. Furthermore we will look at the $\phi_k(x)$ themselves and compare our analytical results with numerical solutions. The results will be encouraging: the approximation of $\phi_k(x)$ up to second order in $k$ is actually quite close to the numerically exact $\phi_k(x)$ even for $k$ at which $\lambda(k)$ is maximal.

\subsection{Growth rate to order $k^{3}$}
Before we completely abandon orders $\left(k/\kappa\right)^{n>2}$, we can harvest one last third-order result, by using a bit of formally fourth-order analysis to obtain the third-order term in the growth rate $\lambda_3$. This will allow us to see how much accuracy in $\lambda$ we are missing by stopping at second order in $k/\kappa$. 

Again by combining (\ref{BdG1}) with its complex conjugate we can eliminate $\varphi_4(x)$ itself from the fourth-order equation and obtain the integral constraint which fixes $\lambda_3$ in terms of functions that we already know:
\begin{align*}
\kappa^2\int \mathrm{sech}^2(\kappa x)\big(2\lambda_1&\mathrm{Im}(\varphi_{3})+2\lambda_2\mathrm{Im}(\varphi_{2})\nonumber\\
+2\lambda_3&\mathrm{Im}(\varphi_{1})+\mathrm{Re}(\varphi_{2})\big)\mathrm{dx}=0\;.
\end{align*} 
The still unknown parameter $B_3$ again makes no contribution to the integral and so we find
\begin{align}\label{lambda3}
\lambda_3=\frac{5\kappa^4-8}{48\sqrt{3}\kappa^2}
\end{align}
In Appendix C this result is compared to approximations for $\lambda(k)$, beyond the linear result of \cite{Kuznetsov}, that have been present previously \cite{KP2000,KP2008}. Here we instead compare our analytical results for $\lambda_{n<4}$ to numerically computed $\lambda(k)$ for various values of $\beta$.

\subsection{Assessment of errors}\label{errors}
\begin{figure*}[htbp]
	\centering
	\includegraphics[width=0.6\textwidth]{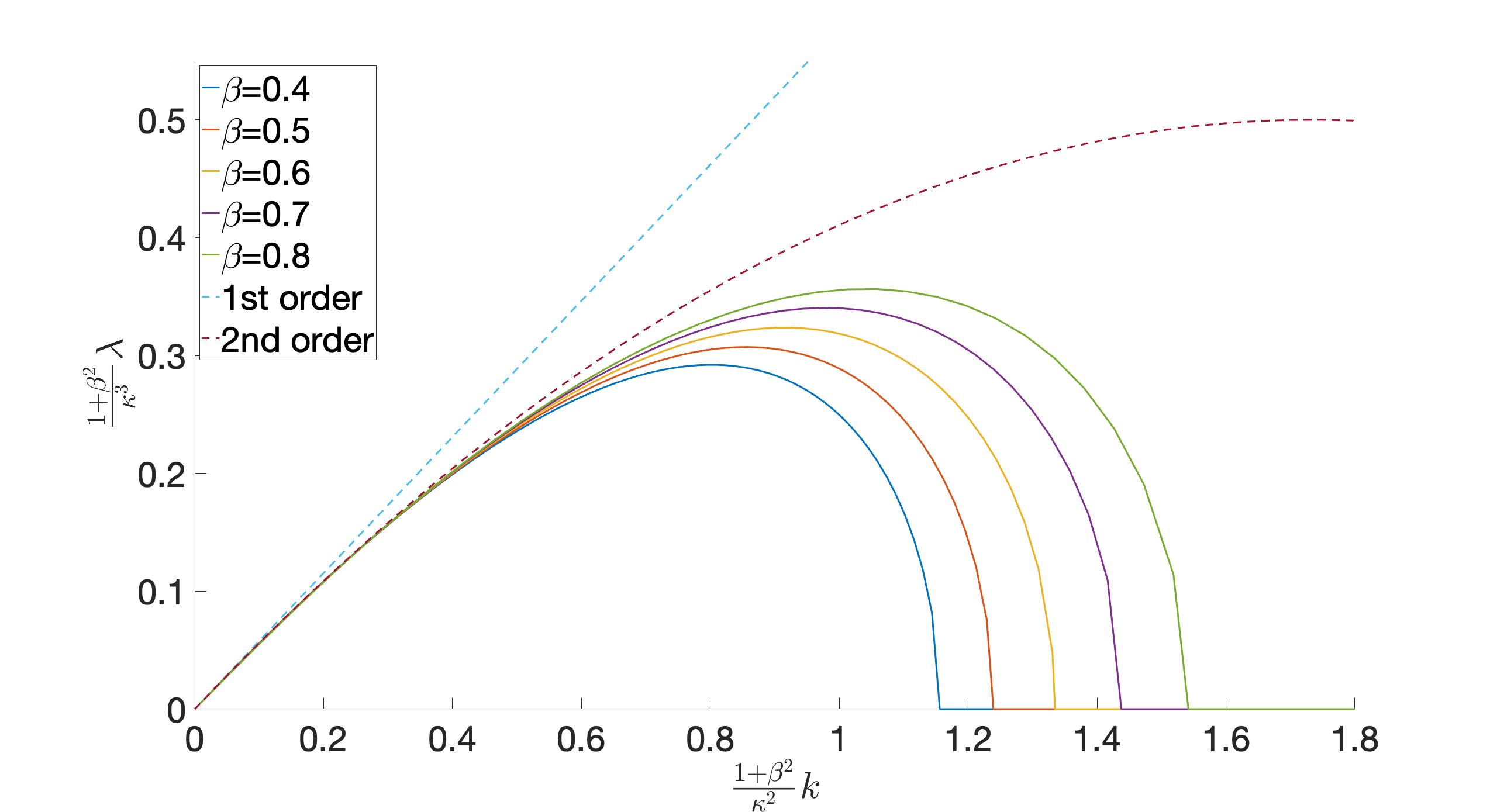}	
	\caption{\textbf{BdG instability growth rates $\lambda(k)$.} This plot shows the growth rates $\lambda(k)$  for finite $k$, for different values of $\beta$, scaled depending on $\beta$ in order to show agreement with analytical results. The dashed lines represent the first and second order approximation}
	\label{NumBdG}
\end{figure*}

As a numerical check on our analytical result, in Fig.~\ref{NumBdG} we show the growth rates $\lambda(k)$ for a range of different $\beta$ values, as computed by discretizing the BdG equations (\ref{BdG1}) into a 1024-by-1024 matrix and numerically finding the positive imaginary eigenvalue. Although until now we have focused on the instability growth rate as a function of $k$ for fixed background soliton grayness parameter $\beta$, $\lambda(k)$ is really $\lambda(k,\beta)$. The first- and second-order approximations $\lambda\doteq \kappa\lambda_1k$ and $\lambda\doteq \kappa\lambda_1k + \lambda_2k^{2}$, however, can be rendered as single curves for all $\beta$ by scaling the axes of the graph in dependence on $\beta$. Fig.~\ref{NumBdG} therefore plots $\left((1+\beta^2)/\kappa^3 \right) \lambda$ versus $\left((1+\beta^2)/\kappa^2\right) k$, so that for all $\beta$ the linear and quadratic approximations according to (\ref{lambda1}) and (\ref{lambda2}) are the single blue and red dashed curves shown in the Figure. Rescaling two axes is not enough to make the full $\lambda(k,\beta)$ into a single curve for all $\beta$, and so the Figure shows the (numerically) exact $\lambda(k,\beta)$ for a finite set of $\beta$ values.

First of all we see that our analytical results are confirmed, inasmuch as all the numerical curves in Fig.~\ref{NumBdG} converge onto the linear approximation for $k\lesssim 0.2 (1-\beta^{2})/(1+\beta^{2})$, and they are all close to the quadratic approximation for $k\lesssim 0.5 (1-\beta^{2})/(1+\beta^{2})$. It is furthermore encouraging to note that the exact curves do not depart dramatically from the quadratic approximation until after their maxima. This means that snake modes with wavelengths long enough for the second-order approximation to be accurate are not much slower-growing than the fastest-growing modes, and so it is plausible that they are at least approximately representative of the instabilities which might be seen in experiments. In contrast the linear approximation really only works for modes which are considerably slower-growing than the fastest modes, and so the effort of extending the results of \cite{Kuznetsov} has been worthwhile.

As a check on our numerical curves, on the other hand, we can see that they all show $\lambda(k)$ going to zero at some finite $k=k_{\text{max}}(\beta)$, implying that snaking of the soliton on wavelengths shorter than some minimum length is no longer an instability.
An analytical formula for $k_{\text{max}}(\beta)$ is given in \cite{Kuznetsov} without derivation; in our notation it reads 
\begin{equation}\label{kmaxbeta}
k_{\text{max}}(\beta)=\sqrt{2\sqrt{1-\beta^{2}+\beta^{4}}-(1+\beta^{2})}\;. 
\end{equation}
In Appendix B we supply a derivation of this result. All the numerical curves in Fig.~\ref{NumBdG} are consistent with it.

\begin{figure*}[hbtp]
	\subfloat[$\beta=0.4$]{
		\includegraphics[width=0.45\textwidth,trim={5.5cm 0cm 8.3cm 3.5cm},clip
		]{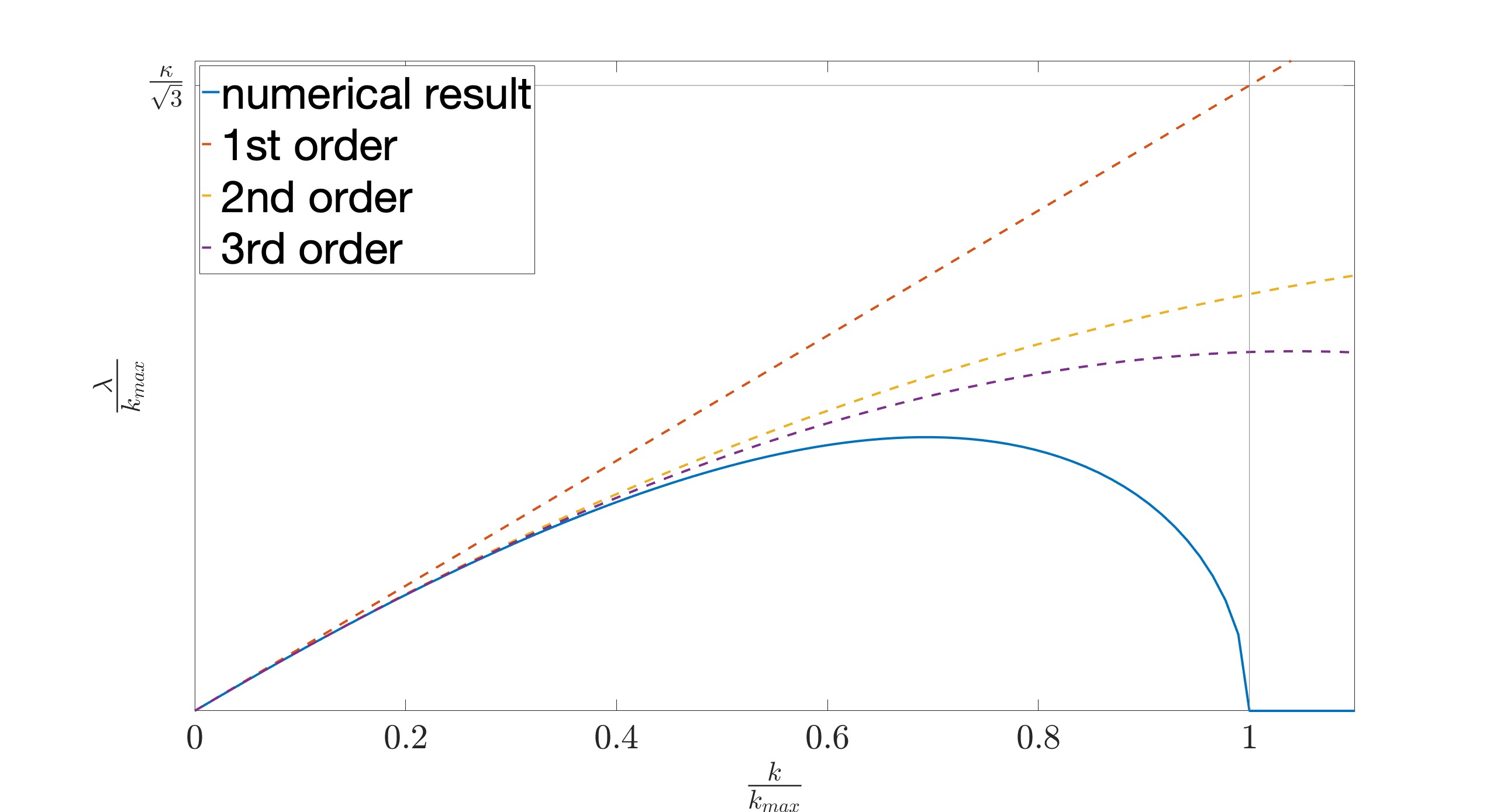}}	
	\label{klambdaplot0}
	\subfloat[$\beta=0.6$]{
		\includegraphics[width=0.45\textwidth,trim={5.5cm 0cm 8.3cm 3.5cm},clip
		]{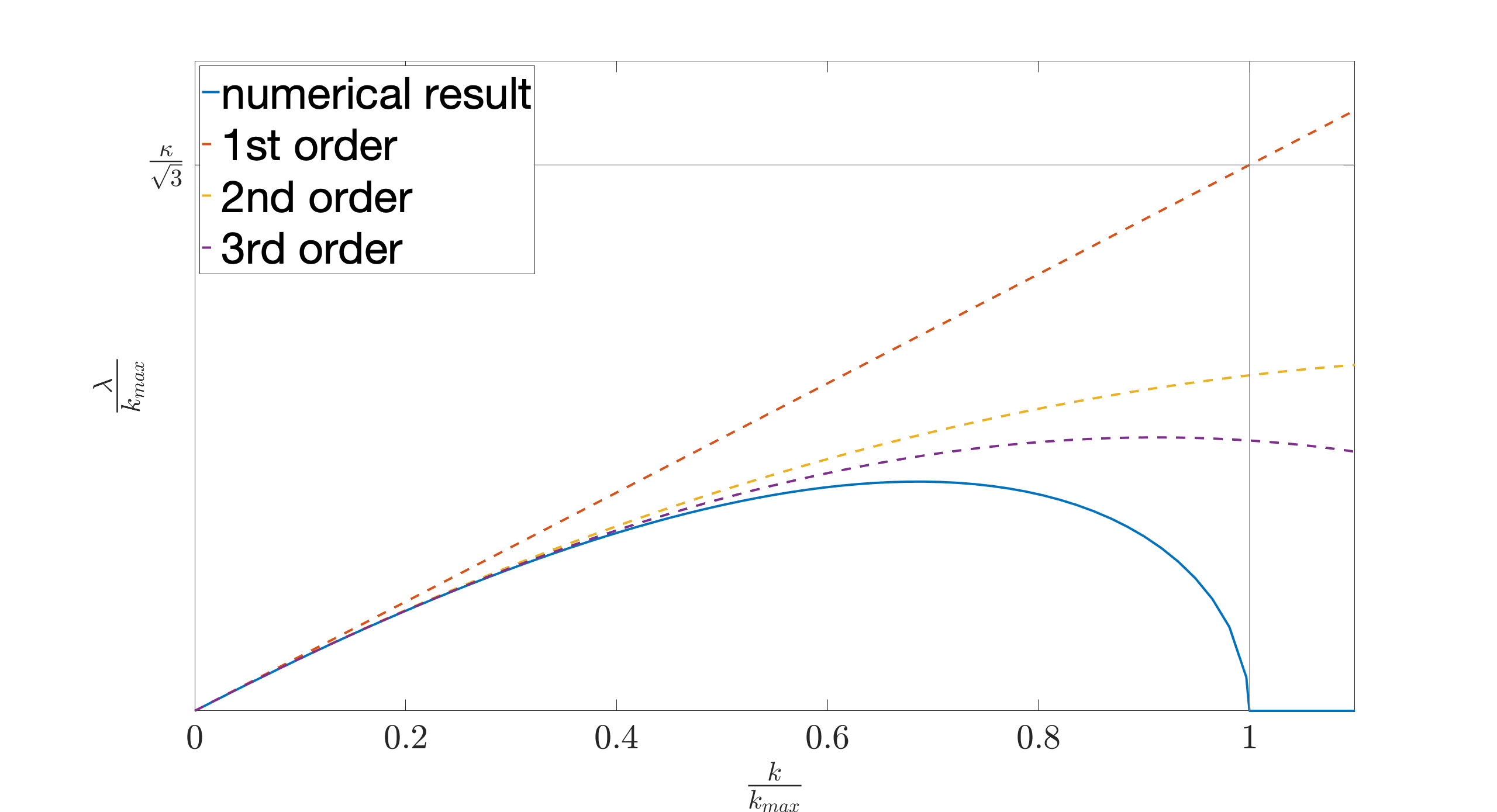}}
	\label{klambdaplot1}
	\subfloat[$\beta=0.8$]{
		\includegraphics[width=0.45\textwidth,trim={5.5cm 0cm 8.3cm 3.5cm},clip
		]{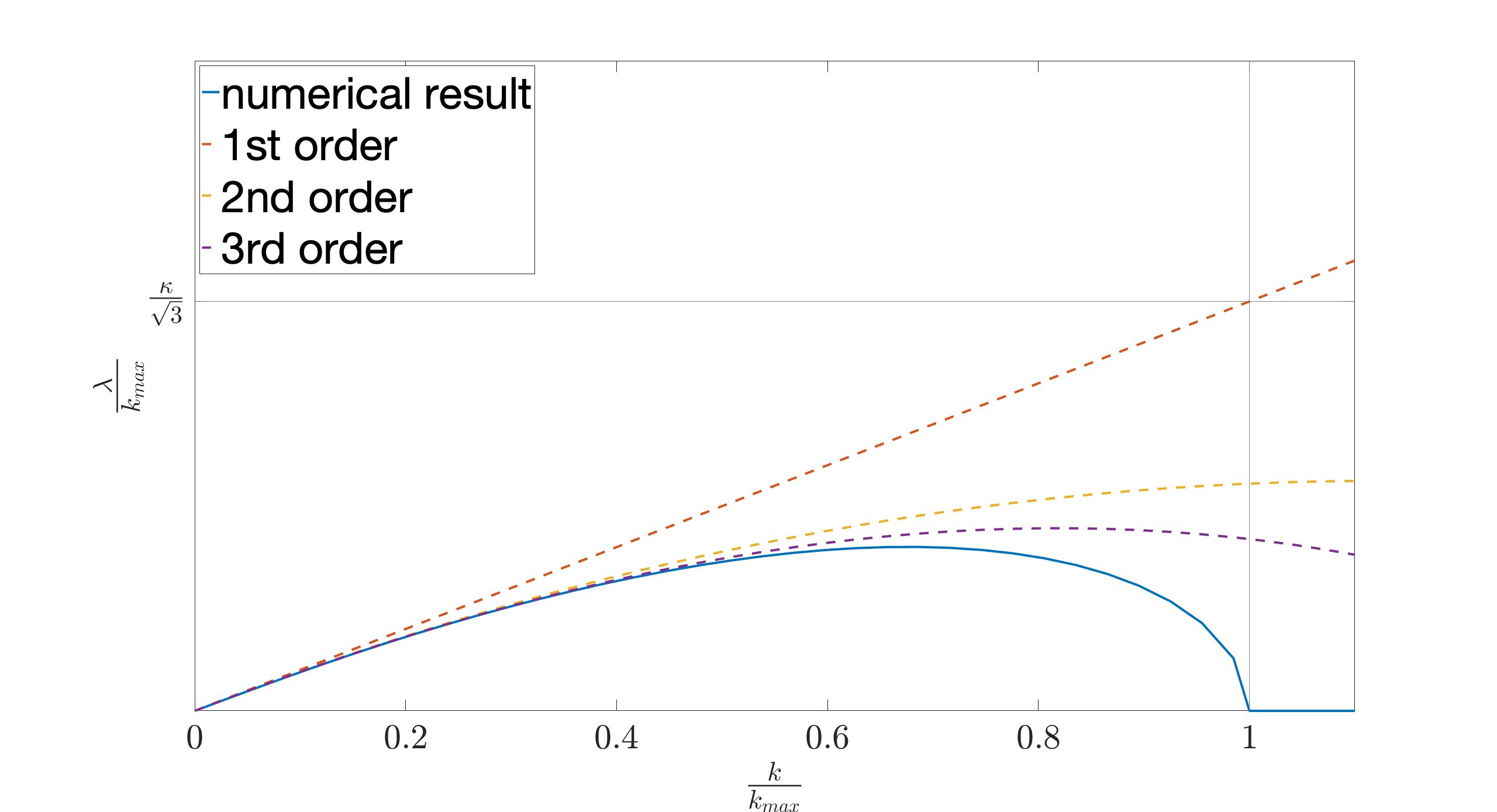}}
	\label{klambdaplot2}
	\caption{\textbf{Growth rate $\lambda/k_{\text{max}}$ vs $k/k_{\text{max}}$ plot for different values of $\beta$.} The plot shows the BdG instability growth rate $\lambda(k)$ normalized by the maximal value $k_{\text{max}}$ for three values of $\beta$, namely $\beta=0.4$, $0.6$ and $0.8$, respectively. The numerical result is represented by the solid line. Dashed lines represent the first, second and third order of approximation.}
	\label{klambdaplot}	
\end{figure*}

To see how much further advantage might be gained from a third-order approximation, as well as to show our first- and second-order approximations on differently scaled axes, Fig.~\ref{klambdaplot} shows three different plots of $\lambda/k_{\text{max}}$ versus $k/k_{\text{max}}$, where $k_{\text{max}}$ is given by (\ref{kmaxbeta}), for $\beta=0.4,0.6,0.8$. (There is no way to rescale the axes to make all the third-order curves coincide.) These plots confirm again that the second order approximation is quite good, even for $\lambda$ quite near its maximum, especially for larger $\beta$. The improvement over the first-order approximation is significant. The additional improvement from the third order, however, is smaller. It is probably not a great enough improvement to justify the considerably greater effort of dealing with the more complicated third-order results.  

\subsection{Comparison between numerical and analytical~$\phi_k(x)$}
\begin{figure*}[hbtp]
	\centering
		\subfloat[$k=0.0404$]{
			\includegraphics[width=0.67\textwidth,trim={0cm 0cm 0cm 0.4cm},clip
			]{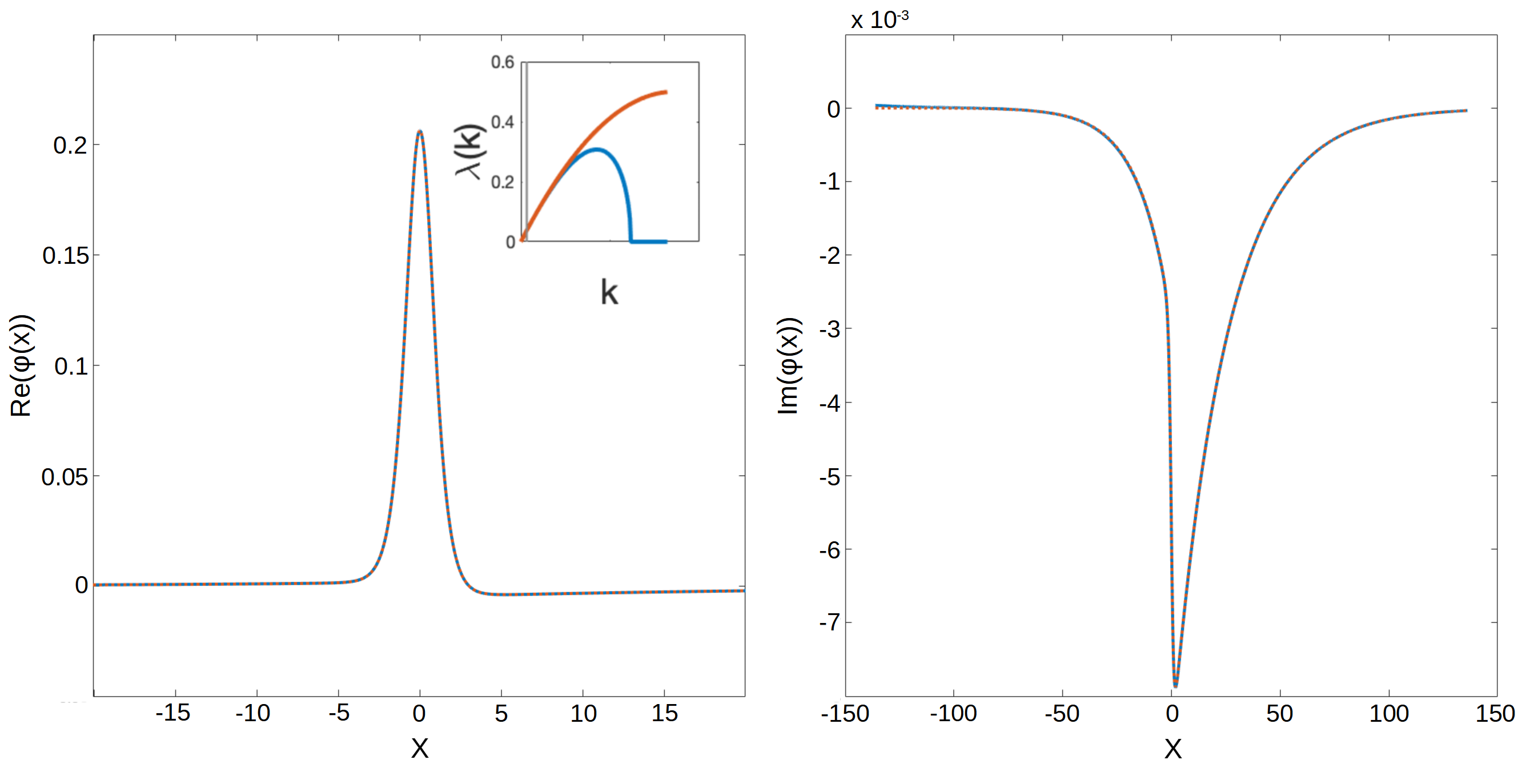}}\label{ComparisonSmallk}
		
		\subfloat[$k=0.1919$]{
			\includegraphics[width=0.67\textwidth,trim={0cm 0cm 0cm 0.4cm},clip
			]{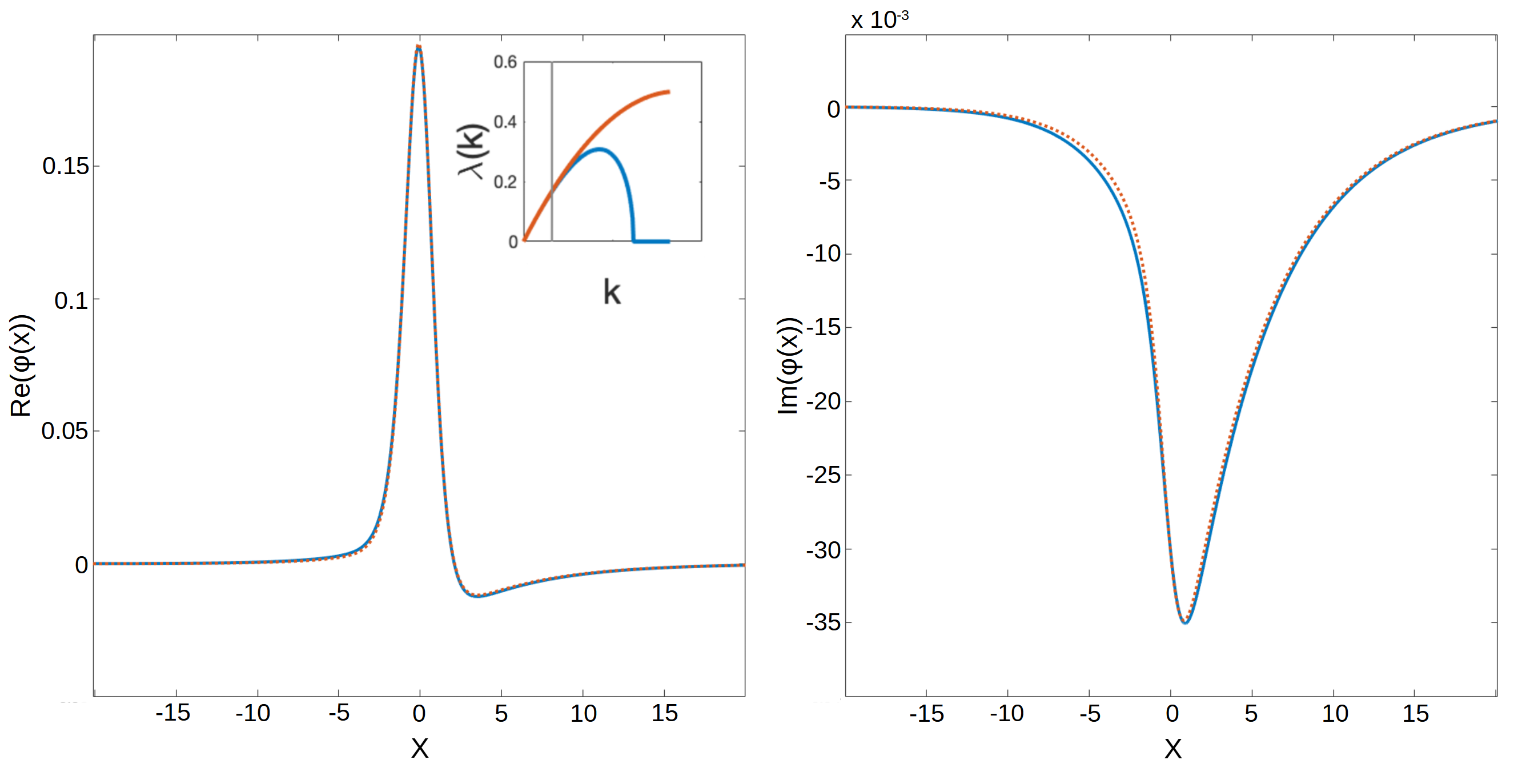}}
		\label{ComparisonMediumk}
		\subfloat[$k=0.5253$]{
			\includegraphics[width=0.67\textwidth,trim={0cm 0cm 0cm 0.4cm},clip
			]{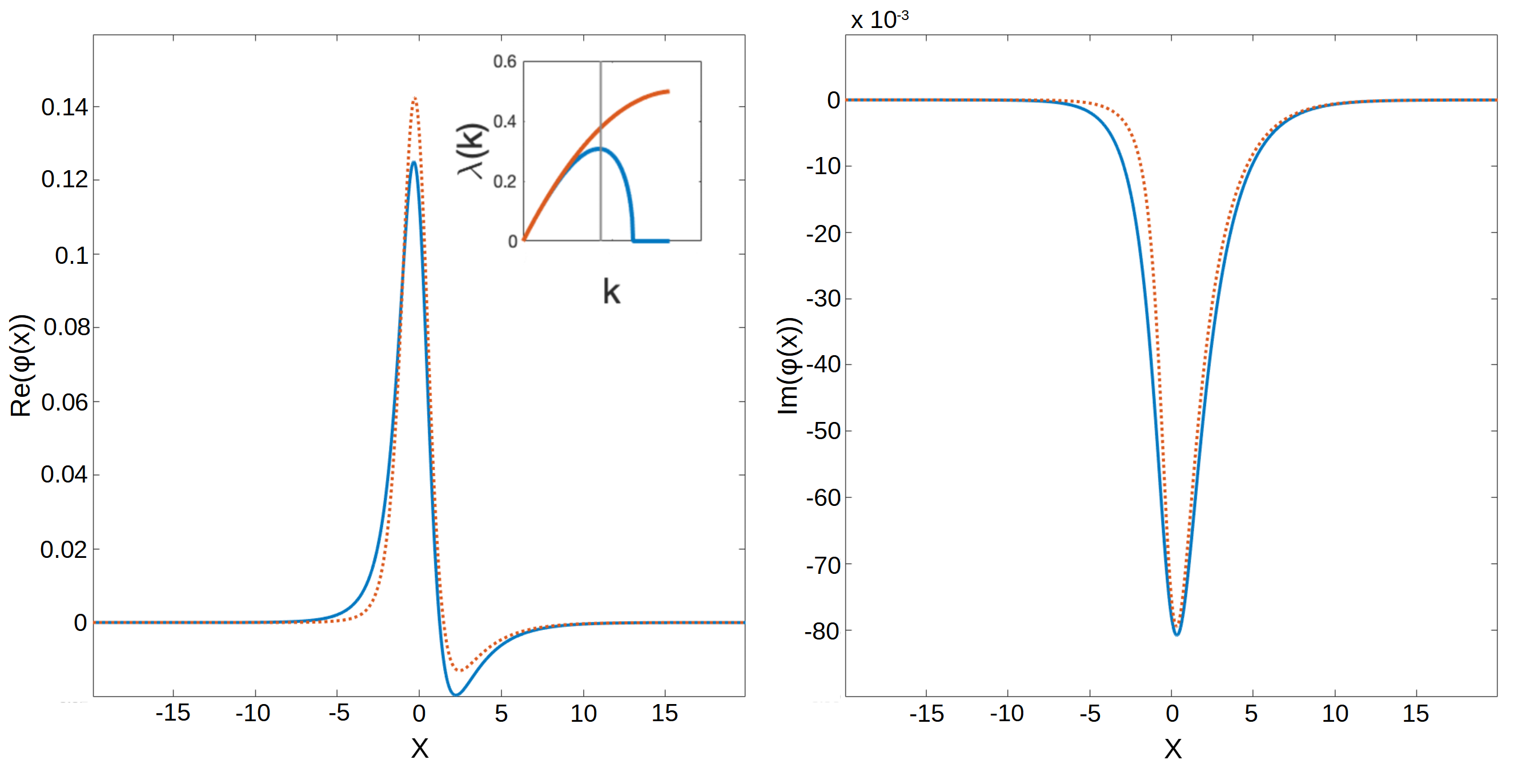}}
		\label{ComparisonLargek}
		\caption{\textbf{Comparison of $\phi_k(x)$} The plot shows pairwise the real (on the left) and the imaginary part (on the right) of $\phi_k(x)$, for the indicated different values of $k$. In all cases the grayness parameter of the soliton is $\beta=0.5$. The solid blue line shows the numerical result and the dotted red line the analytical approximation up to second order in $k/\kappa$. All solutions go to zero for large of $\vert x\vert$, but at $k$-dependent rates, and so optimizing the horizontal range to show most detail has made the horizontal range in the right panel of (a) wider than for the other five panels. The insets in the real-part plots show the numerical and analytically approximate curves of $\lambda(k)$, with a vertical gray line marking the position of $k$ for this panel. The maximum $k$ for which the snake mode is an instability at this soliton grayness is $k_{\text{max}}(\beta=0.5)=0.7435$.}
		\label{phi_comparison}
\end{figure*}

As the final and most important test of our analytical approximations, in this Section we now compare our analytical approximate global solution for $\phi_k(x)$ up to second order in $k/\kappa$ with numerically exact solutions. Our global solutions for $\phi_k(x)$ are constructed from our asymptotically matched inner and outer zone solutions through the envelope procedure (\ref{trafo}, \ref{trafo2}) that was described in subsection \ref{global_k}. Fig. \ref{phi_comparison} shows this comparison between our analytical and numerical results for $\phi_k(x)$ for the ``intermediately gray'' case $\beta=0.5$ (as a generic case), for three illustrative values of $k$. As expected, for small $k$ the agreement between numerical and analytical solution is almost exact. With growing $k$ some discrepancy between the exact and approximate solutions begins to appear, but is small enough to confirm that our higher-order corrections are indeed accurate. 

More noteworthy are the bottom panels in Fig.~\ref{phi_comparison}, which show that even at the $k$ with the highest growth rate, when the numerical and analytical $\lambda(k)$ differ substantially, our second-order approximation for $\phi_k(x)$ remains quite good. Errors are noticeable but still quite small even for the fastest-growing instabilities.      

\section{Conclusions}
The early stages of the snake instability of gray solitons in higher dimensional dilute BECs can be described in mean field theory using the Bogoliubov-de Gennes linearization of the Gross-Pitaevskii equation. Using the method of matched asymptotics, with a modified envelope-factor method of patching together inner and outer solutions, we have found analytical approximations for the mode functions $\phi_k(x)$ as smooth global solutions, and for the growth rates $\lambda(k)$ of snaking with transverse wave number $k$, up to second and third order in $k/\kappa$:
\begin{equation}
\lambda=\frac{\kappa k}{\sqrt{3}}-\frac{1+\beta^2}{6\kappa}k^2 + \frac{5\kappa^4-8}{48\sqrt{3}\kappa^3}k^3+\mathcal{O}(k^4/\kappa^4)\;.
\end{equation}
 Our main result has been to find that as the snake mode first grows the order parameter remains close to a gray soliton in form, with a `snaking' sinusoidal $y$-dependence in its parameters, plus a long-range sound wave dressing:
\begin{widetext}
\begin{equation}
\Psi(x,y,t) = e^{i\beta x}e^{i\theta_C(x,y,t)}e^{-iS(x,y,t)}\tilde\rho(x,y,t)\left(\tilde{\kappa}\tanh\big(\tilde{\kappa}(x-Q)\big) -i\tilde{\beta}\right)+\Phi_\mathrm{NP}(x,y,t)+\mathcal{O}(A_0^2)\;,
\end{equation}
\end{widetext}
where $Q(y,t)=A_0 e^{\lambda t} \cos(ky-\delta)$ and $C$, $S$, $\tilde{\rho}$ and $\tilde{\beta}$ all differ from their initial values in the straight extended soliton by functions proportional to $Q(y,t)$. The specific forms of all these parameters and functions were given above in (\ref{param3}), up to corrections of order $(k/\kappa)^3$. Also in (\ref{param3}) is the explicit form for the non-parametric part of the snake perturbation $\Phi_{NP}(x,y,t)$, which at least at long snaking wavelengths is everywhere a much smaller change in $\Psi$ than the parametric parts of the snake mode.
	
The good news is that the linearized snake mode remains essentially a parametric instability of the gray soliton up to second order in $k/\kappa$. Our results indicate that the second-order approximation is probably worth using as a basis for future quantum mechanical studies of the snake instability based on quantization of parameters in a variational Ansatz as collective coordinates. Although our final expression is lengthy, in a variational calculation all its terms will simply be integrated once to produce the effective Lagrangian or Hamiltonian for the collective coordinates, and our remarkably accurate approximation to the exact Bogoliubov-de Gennes modes will ensure that this effective theory will not have overlooked any qualitatively important behavior, at least in the early stages of the snake instability. The improvement in accuracy with our second-order result over the simpler first-order approximation is significant. In contrast the small further improvement from going to third order is probably not worth the much greater additional effort.

\newpage
\appendix
\onecolumngrid
\section{Bogoliubov-de Gennes solutions in the inner zone}\label{App_A}
In this Appendix we explain how the general BdG solutions for $\varphi_{n}(x)$ that we present in the main text can be found systematically. We begin by distinguishing the real and imaginary parts of $\phi_k(x)$: 
\begin{align}
	\phi_k(x)=R(x)+iS(x)=\sum_{n=0}^{\infty}\left(\frac{k}{\kappa}\right)^{n}[R_{n}(x)+iS_{n}(x)]\label{pert2}
	\end{align}
for real $R$ and $S$. In terms of $R$ and $S$ the BdG equations (\ref{BdG1}) read
\begin{align}
	&-\frac{1}{2}R''(x)+\kappa^2\big[2-3\,\mathrm{sech}^{2}(\kappa x)\big]R(x)+\beta\big[S'(x)-2\kappa\tanh(\kappa x)S(x)\big]
	=-\lambda S(x)-\frac{k^{2}}{2}R(x)
	\label{JR1a}
\end{align}
\begin{align}
	&-\frac{1}{2}S''(x)+\big[2\beta^{2}-\kappa^2\mathrm{sech}(\kappa x)^{2}\big]S(x)-\beta\big[R'(x)+2\kappa \tanh(\kappa x)R(x)\big]
	=\lambda R(x)-\frac{k^{2}}{2}S(x)\label{JS1a} 
	\end{align}
Our perturbative procedure will exploit the fact that for small $k$ the right-hand sides of these equations contain small parameters, since $\lambda=\kappa\lambda_1 k+\lambda_2k^{2}+\lambda_3k^3/\kappa+\mathcal{O}(k^{4})$. The equations at order $\left(k/\kappa\right)^{n}$ will therefore have $R_{n}$ and $S_{n}$ on the left-hand side, and only components of lower $n$, like $R_{n-1}$ and $S_{n-2}$, on the right-hand side. We will work upward in $n$ from $n=0$, determining each $R_{n}$ and $S_{n}$ successively; hence at each order the right-hand sides of our equations will consist of functions that have already been determined at the previous orders.  In other words, we will effectively be solving inhomogeneous differential equations in which the right-hand sides are previously determined sources.

To recognize this pattern, therefore, we will define the right sides of the equations above as power series in $k/\kappa$ whose coefficients are the sources  $\rho_{n}(x)$ and $\sigma_{n}(x)$ respectively:  
\begin{align}
	-\,\kappa^2\left\lbrace\left(\sum_{m=1}^{n}\lambda_m S_{n-m}(x)\right)+\frac{1}{2}R_{n-2}(x)\right\rbrace=:&\,\rho_{n}(x)\label{rhon}\\
	+\,\kappa^2\left\lbrace\left(\sum_{m=1}^{n}\lambda_m R_{n-m}(x)\right)-\frac{1}{2}S_{n-2}(x)\right\rbrace=:&\,\sigma_{n}(x)\label{sigman}\;,
\end{align}
where terms with negative index are set to zero. Our BdG equations (\ref{JR1a}) and (\ref{JS1a}) then appear as inhomogeneous equations at each order, with sources that are fixed functions determined from lower-order equations:
\begin{align}
	&-\frac{1}{2}R_{n}''(x)+\kappa^2\big[2-3\,\mathrm{sech}^{2}(\kappa x)\big]R_{n}(x)+\beta\big[S_{n}'(x)-2\kappa\tanh(\kappa x)S_{n}(x)\big]
	=\rho_{n}(x)
	\label{JR1}
\end{align}
\begin{align}
	&-\frac{1}{2}S_{n}''(x)+\big[2\beta^{2}-\kappa^2\mathrm{sech}(\kappa x)^{2}\big]S_{n}(x)-\beta\big[R_{n}'(x)+2\kappa \tanh(\kappa x)R_{n}(x)\big]
	=\sigma_{n}(x)\label{JS1} 
	\end{align}
The homogeneous parts of (\ref{JR1}) and (\ref{JS1}), namely the left-hand sides, are then the same BdG equations that one has for the zero modes of a one-dimensional gray soliton. The homogeneous problem has been solved completely in Ref.~\cite{Philip}, including its zero modes, using a supersymmetric mapping. By extending this method, we will be able to solve the inhomogeneous equations with sources $\rho_{n}$ and $\sigma_{n}$, order by order in $k$.

The supersymmetric mapping technique of \cite{Philip} is based on the following two differential operators $\hat{Q}$ and $\hat{Q}^{\dagger}$:
	\begin{align}
	\hat{Q}:=&\frac{1}{\sqrt{2}}\left(2\kappa\tanh(\kappa x)+\partial_x\right)\\
	\hat{Q}^{\dagger}:=&\frac{1}{\sqrt{2}}\left(2\kappa\tanh(\kappa x)-\partial_x\right)\;.
	\end{align} 
In terms of these new operators the BdG equations (\ref{JR1}) and (\ref{JS1}) can be re-written exactly as
	\begin{align}
	\rho_{n}(x) &= \hat{Q}^{\dagger}\left[\hat{Q}R_{n}(x)-\sqrt{2}\beta S_{n}(x)\right]\label{JR}\\ 
	\sigma_{n}(x) &= \left[\hat{Q}\hat{Q}^{\dagger}+\left(4\beta^2-2\right)\right]S_{n}(x)-\sqrt{2}\beta\hat{Q}R_{n}(x)\label{JS}
	\end{align}
The algorithmic procedure which generates the $\varphi_{n}(x)$ solutions presented in our main text, as well as the result for $\varphi_{2}(x)$ to which we referred without showing it explicitly, will be to apply the exact homogeneous solutions from \cite{Philip} and to incorporate the sources $\rho_{n}$ and $\sigma_{n}$ by using the associated Green's functions. In fact the general Green's function method can be simplified in this case, by exploiting some convenient properties of $\hat{Q}$ and $\hat{Q}^{\dagger}$.

First of all, the Green's function method for $R(x)$ can be broken down into two simpler stages. We solve for a new source $j_{n}(x)$ such that
	\begin{align}
	\rho_{n}(x)&=:\frac{1}{\sqrt{2}}\hat{Q}^{\dagger}j_{n}(x)\;.\label{jR}
	\end{align}
The reason for doing this is that, once we have found this $j_{n}(x)$, Eqn.~(\ref{JR}) can be reduced to the first-order differential equation
\begin{equation}	
	\hat{Q}R_{n}(x)-\sqrt{2}\beta S_{n}(x)=\frac{1}{\sqrt{2}}j_{n}(x) + \tilde{a}_{n}\cosh^{2}(\kappa x)\label{jR1}
\end{equation}
for any constant $\tilde{a}_{n}$. We solve equation (\ref{jR}) for $j_{n}(x)$ by multiplying it with~$2\,\mathrm{sech}^2(\kappa x)$ and obtaining:
\begin{align}
	-2\,\mathrm{sech}^2(\kappa x)\rho_{n}(x)=\mathrm{\frac{d}{dx}}\left[\mathrm{sech}^2(\kappa x)j_{n}(x)\right]
	\end{align}
with the solution
	\begin{align}
	j_{n}(x)=&\,a_{n}\cosh^{2}(\kappa x)\nonumber\\
	&-2\cosh^2(\kappa x)\int_c^x \mathrm{d\bar{x}}\,\mathrm{sech}^2(\kappa \bar{x})\rho_{n}(\bar{x})\label{jR2}
\end{align} 
where the lower limit of integration $c$ is arbitrary because any change in it merely adds another $\cosh^{2}(\kappa x)$ term that can be absorbed by shifting $a_{n}$. By shifting $a_{n}$ we can also set $\tilde{a}_{n}=0$ in (\ref{jR1}) without loss of generality.
	
	We can now complete the two-stage Green's function solution for $R_{n}(x)$, by multiplying equation (\ref{jR1}) by $\sqrt{2}\cosh^2(\kappa x)$ to obtain
\begin{align}
	\mathrm{\frac{d}{dx}}\left[R_{n}(x)\cosh^2(\kappa x)\right]=\cosh^2(\kappa x)\left[2\beta S_{n}(x)+j_{n}(x)\right]
	\end{align}    
with the immediate solution
	\begin{align}
	&R_{n}(x)=A_{n}\kappa^{2}\,\mathrm{sech}^2(\kappa x)\nonumber\\
	&+\,\mathrm{sech}^2(\kappa x)\int_c^x \mathrm{d\bar{x}}\cosh^2(\kappa\bar{x})\left[2\beta S_{n}(\bar{x})+j_{n}(\bar{x})\right]\label{R}
	\end{align}
for any constant $A_{n}$. Thus, given $\rho_{n}$ and $S_{n}$, we can find $R_{n}$ by integrating in (\ref{R}) after integrating in (\ref{jR2}) to obtain $j_{n}$.

We must now find $S_{n}$, but the equation (\ref{JS}) which determines $S_{n}$ does not reduce so easily to quadratures. It can be simplified, however, by inserting $\hat{Q}R_{n}(x)=\frac{1}{\sqrt{2}}j_{n}(x)+\sqrt{2}\beta S_{n}(x)$ from (\ref{jR1}) with $\tilde{a}_{n}=0$ and finding 
	\begin{align}
	 \sigma_{n}(x)+\beta j_{n}(x)&= \left(\hat{Q}\hat{Q}^{\dagger}-2\kappa^2\right)S_{n}(x)\label{S}\\
	 &\equiv -\frac{1}{2}S_{n}''(x)-\kappa^2\mathrm{sech}^{2}(\kappa x)S_{n}(x) \;.\nonumber
	\end{align}
For $\sigma_{n}+\beta j_{n}\to0$ we have the general homogeneous solution $S_{n}(x)=B_{n}\tanh(\kappa x)+C_{n}[\kappa x\tanh(\kappa x)-1]$. The Green's function may be constructed from these two solutions in the usual way, but in fact we will be able to obtain the solutions we need for $\varphi_{n}$ up to $n=2$ by inspection.

\subsubsection{Zeroth order}
	At order $n=0$ we have $\rho_{0}=\sigma_{0}=0$. In general we could still allow $j_{0}=a_{0}\cosh^{2}(\kappa x)$ for any $a_{0}$, but even without explicitly computing the particular solution that this $j_{0}$ would generate for $S_{0}$, it is easy to see just from inspection of (\ref{S}) at large $|x|$ that a $j_{0}\propto \cosh^{2}(\kappa x)$ could only produce an exponentially growing term $\sim e^{+2\kappa |x|}$ in $S_{0}$. Such exponentially growing functions are indeed among the four linearly independent solutions to the fourth-order BdG equations. We do not need to look closely at our outer zone solutions, however, to recall that they have no exponentially growing terms like $e^{+2\kappa|x|}$. The additional BdG solution which is proportional to $a_{0}$ is therefore one which is ruled out by our boundary conditions, and so we must set $a_{0}=0$.
	
	With $j_{0}=0$, then, we obtain
	\begin{align}
	S_0(x)=B_{0}\tanh(\kappa x)+C_0\left(\kappa x\tanh(\kappa x)-1\right) \;.\label{S01}
	\end{align} 
Inserting this into (\ref{R}) then yields
	\begin{align*}
	&R_0(x)=A_0\,\mathrm{sech}^2(\kappa x)+\frac{\beta}{\kappa}\left[B_0+C_0\left(\kappa x\left(1-\frac{3}{2}\,\mathrm{sech}^2(\kappa x)\right)-\frac{3}{2}\tanh(\kappa x)\right)\right]
	\end{align*} 
	and $\varphi_{0}=R_{0}+iS_{0}$ is the result used in our main text. We found there that matching with the outer zone at $n=0$ required $C_{0}=0$.
		
\subsubsection{First order}
From our zeroth-order result, according to  (\ref{rhon}) and (\ref{sigman}) we have
\begin{align}
\rho_{1}(x)&=-\kappa^2\lambda_1 B_0\tanh(\kappa x)\\
\sigma_{1}(x)&=\lambda_1\beta \kappa B_0+\kappa^2\lambda_1 A_0\,\mathrm{sech}^{2}(\kappa x)\;.
\end{align}
Integrating in (\ref{jR2}) then yields
\begin{align}
j_{1}(x)=-B_0\kappa\lambda_1  + a_1\cosh^{2}(\kappa x)\;,
\end{align}
but any $a_1\not=0$ will again find only an exponentially growing solution for $S_{1}$, which will be ruled out by matching, so we set $a_{1}=0$.
Our equation (\ref{S}) for $S_1(x)$ thus reads
\begin{align}
-\frac{1}{2}S_1''(x)-\kappa^2\mathrm{sech}^2(x)S_1(x)&=\kappa^2\lambda_1 A_0\,\mathrm{sech}^{2}(\kappa x)\;.
\end{align} 
We already know the general homogeneous solution, and the particular solution can obviously be simply a constant, namely $\lambda_1 A_0$, so we have the full general solution
\begin{align}
S_1(x)=B_1\tanh(\kappa x)+C_1\left(\kappa x\tanh(\kappa x)-1\right)-\lambda_1 A_0\;.
\end{align}
Integrating in (\ref{R}) then gives
\begin{align}
R_1(x)=A_1\,\mathrm{sech}^2(\kappa x)+\frac{\beta}{\kappa}\bigg[B_1+C_1\kappa x-\left(\frac{3}{2}C_1+\lambda_1 \left(A_0+\frac{\kappa B_0}{2\beta}\right)\right)\left(\kappa x\,\mathrm{sech}^{2}(\kappa x)+\tanh(\kappa x)\right)\bigg]\;.
\end{align}
Again $\varphi_{1}=R_{1}+iS_{1}$ is the result used in our main text, where we found that matching with the outer zone implied $C_{1}=B_0=0$ but left $A_{1}$ and $B_{1}$ undetermined. As argued in the Section III.E, we can set $A_1$ to zero without loss of generality.
	
\subsubsection{Second order}
Our solutions at order $n=1$ imply
\begin{align*}
\rho_{2}(x) &=\kappa^2\left[A_0\left(\lambda_1^2-\frac{1}{2}\mathrm{sech}^2(\kappa x)\right)-\lambda_1 B_1\tanh(\kappa x)\right]\\
\sigma_{2}(x)& =\kappa^2\bigg[A_0\lambda_2\mathrm{sech}^2(\kappa x)+\lambda_1\frac{\beta}{\kappa}\left[B_1-\lambda_1 A_0 \left(\kappa x\:\mathrm{sech}^2(\kappa x)+\tanh(\kappa x)\right)\right]\bigg] 
\end{align*}
Integrating in (\ref{jR2}) then determines
\begin{align*}	
j_{2}(x) =\sinh(\kappa x)\cosh(\kappa x)\left[\frac{2}{3}A_0\kappa\left(1-3\lambda_1^2\right)\right]+\frac{A_0}{3}\kappa\tanh(\kappa x)-B_1\kappa\lambda_1 + a_{2}\cosh^{2}(\kappa x)\;.
\end{align*}
To avoid unmatchable growing terms in $S_{2}$ we must set $a_{2}=0$. Since $\sinh(\kappa x)\cosh(\kappa x)=\sinh(2\kappa x)$ is another exponentially growing source, independent of $\cosh^{2}(\kappa x)$ because it has opposite parity, we must now also set $1-3\lambda_1^2=0$, as we established in the main text by a different calculation which was also based on the fact that $S_{2}$ and $R_{2}$ cannot be exponentially growing.
	
The differential equation (\ref{S}) for $S_2(x)$ now reads
\begin{align}
-&\frac{1}{2}S_2(x)''-\kappa^2\mathrm{sech}^2(\kappa x)S_2(x)=\kappa^2\left[\lambda_2A_0\mathrm{sech}^2(\kappa x)-A_0\frac{\beta}{\kappa}\lambda_1^2\kappa x\:\mathrm{sech}^2(\kappa x)\right]\;.
\end{align}
Either by inspection or with the Green's function we can then find the general solution
\begin{align}
S_2(x) =& B_2\tanh(\kappa x)+C_2\left(\kappa x\tanh(\kappa x)-1\right)- \lambda_2A_0+A_0\beta\lambda_1^2 x\;.
\end{align}
Solving the integral in (\ref{R}) we finally find $R_2(x)$ to be
\begin{align}
R_2(x) =& A_2\,\mathrm{sech}^2(\kappa x)-\left[\frac{\beta}{\kappa}\lambda_2A_0+\frac{\lambda_1B_1}{2}+\frac{3}{2}\frac{\beta}{\kappa} C_2\right]\left[\tanh(\kappa x)+\kappa x\:\mathrm{sech}^2(\kappa x)\right]+C_2\beta x\nonumber\\
&+\frac{1}{2}\lambda_1^2\frac{\beta^2}{\kappa^2} A_0\left[\kappa^2 x^2\mathrm{sech}^2(\kappa x)+2\kappa x\tanh(\kappa x)\right]+\frac{A_0}{6}+\frac{\beta}{\kappa} B_2 - \frac{1}{2}\lambda_1^2\frac{\beta^2}{\kappa^2} A_0\;.
\end{align} 
Again the expression used in Section IV corresponds to $\varphi_{2}=R_2+iS_2$. The final result used in the main text is found by matching the only term in $R_{2}(x)+iS_{2}(x)$ which is proportional to $|x|$ at large $|x|$, namely $C_2\beta\pm\lambda_1^2\beta^2A_0/\kappa+i\left(\pm C_2\kappa+\lambda_1^2\beta A_0\right)$, with the corresponding $+|x|$ term for the outer zone solutions in (\ref{Outer2}). This gives
\begin{align}
B_{1}&= -\frac{2\beta\lambda_1^2}{\sqrt{\lambda_1^2+1}}A_0\\
C_2&=\frac{\lambda_1}{\kappa}\left(\beta B_1+\sqrt{\lambda_1^2+1}\,A_0\right)\;.
\end{align}
To fully obtain the solution up to second order, we need to determine $B_2$, which will be fixed by extending the calculation to third order and again performing the matching.

\subsubsection{Third order}
Taking into account our previous solutions the equations (\ref{rhon}) and (\ref{sigman}) yield for $n=3$:
\begin{align*}
\rho_3&=-\kappa^2\left(S_2(x)\lambda_1+S_1(x)\lambda_2+\frac{R_1(x)}{2}\right)\\
\sigma_{3}&=\kappa^2\left(R_2(x)\lambda_1+R_1(x)\lambda_2+R_0(x)\lambda_3-\frac{S_1(x)}{2}\right)
\end{align*}   
In the same way as was done for the second order calculation we find $j_3(x)$ by applying (\ref{jR2}) and inserting all already known quantities, to obtain
\begin{align}
j_3(x)=\frac{A_0\left(3\beta\kappa\lambda_2+\beta^2\kappa x-2\kappa x + \beta\right)-A_0\left(\beta^2+6\kappa\lambda_2+1\right)\sinh(2\kappa x)-A_0\beta\kappa x \tanh(\kappa x)-3 \kappa B_2}{3\sqrt{3}}+a_3 \cosh^2(\kappa x)\;.
\end{align}
To eliminate all terms in $S_3$ that cannot be matched to the outer zone solution because they are exponentially growing, we must set $a_3=0$ (to eliminate the even function $\cosh^2(\kappa x)$) and also fix $\left(\beta^2+6\kappa\lambda_2+1\right)=0$ (to eliminate the odd function $\sinh(2\kappa x)$). This provides the main text result $\lambda_2=-(1+\beta^2)/6\kappa$.   

The differential equation (\ref{S}) for $S_3(x)$ now reads
\begin{align}
-\frac{1}{2}S_3(x)''-\kappa^2\mathrm{sech}^2(\kappa x)S_3(x)=& \left(\frac{A_0 \kappa ^2\beta^2}{6 \sqrt{3}}\right) x^2\,\text{sech}^2(\kappa  x)+\left(\frac{A_0 \beta  \kappa\left(1-4\kappa^2\right) }{6 \sqrt{3}}\right) x\,\text{sech}^2(\kappa  x)\\
&+\left(\frac{A_0 \beta^3 }{6 \sqrt{3}}\right) \tanh (\kappa  x)+A_0 \kappa ^2 \lambda_3\,\text{sech}^2(\kappa  x)+\frac{A_0(1+3\kappa^2)}{6 \sqrt{3}}\;.
\end{align}
The general solution can again be found with the Green's function or by inspection: 
\begin{align}\label{S3}
S_3(x)=&\,B_3\tanh(\kappa x)+C_3\left(\kappa x \tanh(\kappa x)-1\right)+ \left(\frac{A_0 \kappa ^2\beta^2}{6 \sqrt{3}}\right)\mathrm{P}(x)-\left(\frac{A_0 \beta  \kappa\left(1-4\kappa^2\right) }{6 \sqrt{3}\kappa^2}\right)x\nonumber\\
&+\left(\frac{A_0 \beta^3 }{6 \sqrt{3}}\right) \left(x^2\tanh(\kappa x)-\frac{2 x}{\kappa}\right)-A_0 \lambda_3+\frac{A_0(1+3\kappa^2)}{6 \sqrt{3}}\left(x^2+\kappa^2\mathrm{P}(x)\right)\;.
\end{align}
where $\mathrm{P}(x)$ is an expression containing the dilogarithm function $\mathrm{Li}_2(x)$:
\begin{align}
\mathrm{P}(x)=\frac{1}{\kappa^4}\Bigg\lbrace\left[\left((\mathrm{Li}_2\left(-e^{-2\kappa x}\right)+\frac{\pi^2}{12})\left(1-e^{-2\kappa x}\right)-2\left(\kappa^2x^2+2\kappa x\right)\right)\left(1+e^{-2\kappa x}\right)^{-1}\right]-2\,\mathrm{ln}(1+e^{-2\kappa x})\Bigg\rbrace
\end{align}
Inserting this $S_3(x)$ result into the integral in (\ref{R}) yields the following $R_3(x)$:
\begin{align}\label{R3}
R_3(x)=-\frac{1}{256\kappa^3}\Bigg\lbrace &p_1(x)\,\mathrm{sech}^2(\kappa x)+\left(c_1\,\mathrm{Li}_2(-e^{2\kappa x})-c_2\,\mathrm{Li}_2(-e^{-2\kappa x})\right)\,\mathrm{sech}^2(\kappa x)+p_2(x)\tanh(\kappa x)\nonumber\\
+&c_3\ln\left(\frac{1+e^{2\kappa x}}{1+e^{-2\kappa x}}\right)x\,\mathrm{sech}^2(\kappa x)-c_4\ln(1+e^{-2\kappa x})\tanh(\kappa x)+2c_2\,\mathrm{Li}_2(-e^{-2\kappa x})+p_3(x)\Bigg\rbrace\;,
\end{align}
where $p_i(x)$ are polynomials in $x$ and $c_i$ are constants. They read as follows:
\begin{align*}
p_1(x)=&-2 \sqrt{3} A_0 \left(\pi ^2 \beta\kappa^2 -3\left(\kappa^4-\beta^2\right)\right)+x \left(36 \kappa ^3 \left(\sqrt{3} B_2 \kappa +9 \beta  C_3\right)-6A_0 \beta  \kappa  \left(\sqrt{3} \kappa ^4+\kappa ^2 \left(22 \sqrt{3}-36 \lambda_3\right)-\sqrt{3}\right)\right)\\
&\quad+4 \sqrt{3} A_0 \beta^3 \kappa ^3 x^3+12 \sqrt{3} A_0 \kappa ^2 x^2 \left(-14 \beta  \kappa ^2+\kappa ^4-\beta^2\right)+108 \beta  B_3 \kappa ^2\\
p_2(x)=&+6 \left(-\sqrt{3}A_0 \beta  \kappa ^4+2 A_0 \beta  \kappa ^2 \left(18 \lambda_3+\sqrt{3}\right)+\sqrt{3} A_0 \beta +6 \kappa ^2 \left(\sqrt{3} B_2 \kappa +9 \beta  C_3\right)\right)+12 \sqrt{3} A_0 \beta^3 \kappa ^2 x^2\\
&\quad+6 x \left(4 \sqrt{3} A_0 \kappa  \left(\kappa ^4-\beta^2\right)-24 \sqrt{3} A_0 \beta  \kappa ^3\right)\\
p_3(x)=&+4 \left(\sqrt{3} A_0 \kappa ^2 \left(\pi ^2 \beta -3 \left(\kappa ^2+1\right)\right)+3 \sqrt{3} A_0-54 \beta B_3 \kappa ^2\right)+4 x \left(3 \sqrt{3} A_0 \beta  \kappa  \left(10 \kappa ^2-1\right)-54 \beta  C_3 \kappa ^3\right)\\
&\quad+12 \sqrt{3} A_0 \kappa  x^2 \left((4 \beta -2) \kappa ^3+\kappa ^5+\kappa \right)\\
c_1=\;&48 \sqrt{3} A_0 \beta  \kappa ^2\\
c_2=\;&24 \sqrt{3} A_0 \beta  \kappa ^2\\
c_3=\;&96 \sqrt{3} A_0 \beta  \kappa ^3 \\
c_4=\;&96 \sqrt{3} A_0 \beta  \kappa ^2
\end{align*} 

The growing complexity of our calculation at third order is still further increased by the need to extend our previous result for the outer zone spatial decay rate $\gamma$ (\ref{gamma}) to higher order in $k/\kappa$, in order to perform our matching of inner and outer zone solutions consistently to order $(k/\kappa)^3$. Finally performing this matching by comparing the part of $\varphi_{3}=R_3(x)+iS_3(x)$ proportional to $\vert x\vert$ at large $\vert x\vert$ with the corresponding term in the outer zone solution we find:
\begin{align}
B_2&=-\frac{11}{24}A_0\beta\kappa\\
C_3&=\frac{4+3\kappa^2+15\kappa^4}{24\sqrt{3}\,\kappa^2}A_0
\end{align}   
By determining $B_2$ we have finally obtained the full solution for $R_k$ and $S_k$ and thereby for $\phi_k$ up to the second order. To obtain the full third order solution we would need to push our calculation to fourth order to determine the still unknown coefficient $B_3$. This poses the new and greater challenge of dealing with the dilogarithm function $\mathrm{Li}_2(-e^{2\kappa x})$.

\section{Upper limit $k_{\text{max}}$ for dynamical instability}
In this Appendix we provide a pedagogical derivation of the formula from Ref.~\cite{Kuznetsov} for the maximum $k$ value $k_{\text{max}}(\beta)$ for which the snake mode is unstable (\textit{i.e.} the growth rate $\lambda$ is real). The starting point for this calculation is the observation that when $\lambda$ changes continuously from purely real to purely imaginary (providing a real frequency $\omega = \pm i\lambda$), it can only do so by passing through zero. 

We therefore begin by setting $\lambda=0$ within the equation (\ref{BdG1}). This allows the known gray soliton zero mode solutions \cite{Philip} for $k=0$, but the definition of $k_{\text{max}}$ is that there will also exist normalizable solutions with $\lambda=0$ for $k=k_{\text{max}}$. As in Appendix \ref{App_A}, we define the real and imaginary part of $\phi_k(x)=R(x)+iS(x)$. We also use the operators $\hat{Q}$ and $\hat{Q}^{\dagger}$ that were introduced in Appendix \ref{App_A}, and so write (\ref{BdG1})
\begin{align}
-\frac{k_{\text{max}}^2}{2}R&=\hat{Q}^{\dagger}\hat{Q}R-\sqrt{2}\beta\hat{Q}^{\dagger}S\label{BB1}\\
-\frac{k_{\text{max}}^2}{2}S&=\left[\hat{Q}\hat{Q}^{\dagger}+(4\beta^2-2)\right]S-\sqrt{2}\beta\hat{Q}R\;.\label{BB2}
\end{align} 

Introducing the ancillary function $f(x)$ through the definition $R=:\hat{Q}^{\dagger}f$, (\ref{BB1}) becomes
\begin{equation}
\hat{Q}^{\dagger}\left(\hat{Q}\hat{Q}^{\dagger}f + \frac{k_{\text{max}}^{2}}{2}-\sqrt{2}\beta S\right) = 0\;.
\end{equation}
Since the only function which is annihilated by $\hat{Q}^{\dagger}=2^{-1/2}(2\kappa\tanh\kappa x -\partial_{x})$ is $\cosh^{2}\kappa x$, 
we must therefore have
\begin{equation}
\hat{Q}\hat{Q}^{\dagger}f + \frac{k_{\text{max}}^{2}}{2}-\sqrt{2}\beta S = Z \cosh^{2}\kappa x
\end{equation}
for some constant $Z$. Simply by considering this equation asymptotically at large $x$, however, it is easy to see that no normalizable solutions can appear unless $Z=0$. We can therefore set $Z=0$, and then use this result to simplify (\ref{BB2}), obtaining as an equivalent pair of equations to (\ref{BB1}, \ref{BB2})
\begin{align}
\hat{Q}\hat{Q}^{\dagger}f&=\sqrt{2}\beta S - \frac{k_{\text{max}}^2}{2}f\label{App_eq_1}\\
\hat{Q}\hat{Q}^{\dagger}S&=\left[2\kappa^2-\frac{k_{\text{max}}^2}{2}\right]S-\frac{k_{\text{max}}^2}{\sqrt{2}}\beta f\;.\label{App_eq_2}
\end{align} 

Equations (\ref{App_eq_1}) and (\ref{App_eq_2}) are a system of coupled linear equations for $f(x)$ and $S(x)$. We can decouple them by considering linear combinations $uf+vS$ for constants $u,v$, since (\ref{App_eq_1}) and (\ref{App_eq_2}) imply
\begin{align}\label{qq}
\hat{Q}\hat{Q}^{\dagger}\left(uf+vS\right)=-\frac{k^2}{2}\left(u+\sqrt{2}\beta v\right)f+\left(\sqrt{2}\beta u-\frac{k^2}{2}v+2\kappa v\right)\;.
\end{align}
The decoupling will succeed, leaving us with the single differential equation
\begin{equation}\label{sepsucceed}
\hat{Q}\hat{Q}^{\dagger}\left(uf+vS\right) = h\left(uf+vS\right)
\end{equation}
for some constant $h$, if we can choose $u$ and $v$ in such a way that
\begin{align*}
-\frac{k_{\text{max}}^2}{2}(u+\sqrt{2}\beta v) &\overset{!}{=}h u\\
\sqrt{2}\beta u-\frac{k_{\text{max}}^2}{2}v+2\kappa^2v &\overset{!}{=}h v \;.
\end{align*}  
In matrix form this separation condition for $u$ and $v$ reads
\begin{align}\label{hmatrix}
h
\begin{pmatrix}
u\\
v
\end{pmatrix}
=
\begin{pmatrix}-\frac{k^2}{2}& -\frac{k^2}{\sqrt{2}}\beta\\
\sqrt{2}\beta& 2\kappa^2-\frac{k^2}{2}
\end{pmatrix}
\begin{pmatrix}
u\\
v
\end{pmatrix}\;.
\end{align}
Possible values of $h$ are therefore simply the eigenvalues of this matrix:
\begin{align}\label{hev}
h=\kappa^2-\frac{k_{\text{max}}^2}{2}\pm\sqrt{\kappa^4-k_{\text{max}}^2\beta^2}\;.
\end{align}
Since we are looking for $k_{\text{max}}>0$, we can see that we must have $h<2\kappa^{2}$.

A second condition on $h$ is also given, however, by Eqn.~(\ref{sepsucceed}): if $R$ and $S$ are to be normalizable then $h$ must be one of the eigenvalues of the operator 
\begin{equation}\label{QQ}
\hat{Q}\hat{Q}^{\dagger}\equiv-\frac{1}{2}\mathrm{\frac{d^2}{dx^2}}-\kappa^2\,\mathrm{sech}^2(\kappa x)+2\kappa^2\;,
\end{equation}
which is the quantum mechanical Hamiltonian for a particle in one dimension subject to a $\mathrm{sech}^{2}$ potential well. Straightforward differentiation will confirm that one eigenfunction of $\hat{Q}\hat{Q}^{\dagger}$ is $\mathrm{sech}(\kappa x)$, having the eigenvalue $3\kappa^{2}/2$, while another eigenfunction is $\tanh(\kappa x)$, with the eigenvalue $2\kappa^{2}$. Since the first of these eigenfunctions has no zeroes, and the second has one zero, by a well-known theorem about the ordering of energy eigenstates these two eigenvalues must be the lowest and second-lowest, respectively. 

We thereby conclude that $h=3\kappa^{2}/2$ is the only possible value that $h$ can have which is less than $2\kappa^{2}$, and so (\ref{hev}) must read
\begin{align*}\label{hev2}
\frac{3}{2}\kappa^{2}=\kappa^2-\frac{k_{\text{max}}^2}{2}\pm\sqrt{\kappa^4-k_{\text{max}}^2\beta^2}\;.
\end{align*}
Solving for $k_{\text{max}}$ while remembering $\kappa^{2}=1-\beta^{2}$ yields the unique positive real root
\begin{align}
k_{\text{max}}=\sqrt{2\sqrt{1-\beta^2+\beta^4}-(1+\beta^2)}	
\end{align}
as reported in Ref.~\cite{Kuznetsov} and used in our main text.

%\begin{widetext}
\section{Comparison with previous analytical formulas for the growth rate}
We can compare our analytical result for the growth rate $\lambda(k)$ with findings in previous publications, in particular the works by Kivshar and Pelinovsky \cite{KP2000} and by Kamchatnov and Pitaevskii \cite{KP2008}. Our final result for $\lambda(k)$ up to the third order in $k$ is given as follows:
	\begin{align}\label{lambda}
	\lambda(k)=\frac{\sqrt{1-\beta^2}}{\sqrt{3}}k-\frac{1+\beta^2}{6\sqrt{1-\beta^2}}k^2 - \frac{3 + 10\beta^2- 5\beta^4}{48\sqrt{3}(1-\beta^2)^{3/2}}k^3+\mathcal{O}(k^4)\;.
	\end{align}
In \cite{KP2000} the authors present as their Eqn.~(3.9) a quadratic equation in $\lambda(k)$ which in our notation would read
\begin{align}\label{KP2000Gamma}
	\lambda^2+\frac{1+\beta^2}{3\sqrt{1-\beta^2}}k^2\,\lambda-\frac{1-\beta^2}{3}k^2=\mathcal{O}(k^4)\:.
\end{align} 
The positive root of this equation yields $\lambda(k)$ up to the same order of corrections $\mathcal{O}(k^4)$ as our (\ref{lambda}) above, but with a different third-order term: 
\begin{align}\label{KPargGsol}
	\lambda_{[4]}(k)= \frac{\sqrt{1-\beta^2}}{\sqrt{3}}k-\frac{1+\beta^2}{6\sqrt{1-\beta^2}}k^2 + \frac{(1 + \beta^2)^2}{24\sqrt{3}(1-\beta^2)^{3/2}}k^3+\mathcal{O}(k^4)\;.
\end{align}
Inspection of our Fig.~4 above confirms that the third-order term in $\lambda(k)$ is in fact negative, as in our solution (\ref{lambda}): our third-order curves in Fig.~4 fall closer to the numerically exact curves by being \emph{lower} than the second-order curves. The result in \cite{KP2000} is thus only correct up to errors of third order in $k$, not up to fourth order corrections as it is presented to be.   
Ref.~\cite{KP2008} also presents a quadratic equation for the growth rate, its Eqn.~(10); in our notation this one reads
\begin{align}
\lambda^2=\frac{1-\beta^2}{3}k^2-\frac{1+\beta^2}{3\sqrt{3}}k^3\;,
\end{align} 
and it is said to be valid for $k\ll k_\mathrm{max}$. Taking the positive square root and Taylor-expanding to third order in $k$ yields
\begin{align}\label{KP2008res}
\lambda_{[6]}(k)\doteq\frac{\sqrt{1-\beta^2}}{\sqrt{3}}k-\frac{1+\beta^2}{6\sqrt{1-\beta^2}}k^2-\frac{(1+\beta^2)^2}{24\sqrt{3}(1-\beta^2)^{3/2}}k^3\;,
\end{align}
which agrees with (\ref{KPargGsol}) except for the sign of its third-order term---and thus has the correct sign for this term--- but still disagrees at third order with our result (\ref{lambda}). 
%\end{widetext}

As noted in our main text, both \cite{KP2000} and \cite{KP2008} give the second-order term in $\lambda(k)$ correctly. Eqn.~(3.9) of \cite{KP2000} and Eqn.~(10) of \cite{KP2008} are both presented without derivation, however, merely citing \cite{Kuznetsov} for the result. The only equation in \cite{Kuznetsov} which is similar to the quadratic equations of \cite{KP2000} and \cite{KP2008}, however, is its Eqn.~(27), which refers \emph{not} to the growth rate of the general gray soliton snake instability, but rather to the growth rate of the similar instability under the Kadomtsev-Petviashvili (KP) equation. In our notation Eqn.~(27) of \cite{Kuznetsov} would read
\begin{equation}\label{KPres}
\lambda_{KP} = \frac{k}{\sqrt{3}}\sqrt{1-\beta^2 - \frac{X}{\sqrt{3}}k}
\end{equation}
for $X=2$. The reason that \cite{Kuznetsov} mentions this KP result (taken with citation from \cite{KuznetsovMusher}) is that the Gross-Pitaevskii equation reduces to the KP equation for long-wavelength perturbations around a gray soliton of small $\kappa = \sqrt{1-\beta^2}$ (shallow amplitude and large width). Refs.~\cite{KP2000} and \cite{KP2008} appear to have taken this KP result (\ref{KPres}) from \cite{Kuznetsov} and reinterpreted it as a result for Gross-Pitaevskii gray solitons of arbitrary $\beta$, by replacing $X$ with different $k$- and/or $\beta$-dependent expressions that reduce to 2 for small $k$ and $\beta^2\to 1$. Unfortunately neither of these procedures appears to be accurate for smaller $\beta$. It may be noted that our result (\ref{lambda}), (\ref{KP2008res}) from \cite{KP2008}, and the KP result (\ref{KPres}) quoted in \cite{Kuznetsov} from \cite{KuznetsovMusher} all agree up to fourth-order corrections when $\beta^2\to1$. 

On the other hand we have to admit that our formally more accurate third-order term in $\lambda(k)$ represents no real improvement on the results of \cite{KP2008}, because \cite{KP2008} has essentially solved the $\lambda(k)$ problem completely for the gray soliton snake mode, by providing an accurate and yet reasonably compact formula for the whole range of wavenumber $k$. This formula is obtained in \cite{KP2008} by multiplying together perturbative expansions around the two ends of the unstable $k$ range. In our notation this formula reads \cite{KP2008}:
\begin{align*}
	&\Gamma^2(k)=f(k)(k_{\text{max}}-k)\quad\text{with}\quad f(k)=a(\kappa)k^2+b(\kappa)k^3+c(\kappa)k^4\\
	&\quad\quad a(\kappa)=\frac{\kappa^2}{3 \, k_{\text{max}}}\\
	&\quad\quad b(\kappa)=\left(\frac{\kappa^2}{k_{\text{max}}}-\frac{2-\kappa^2}{\sqrt{3}}\right)\frac{1}{3 \, k_{\text{max}}}\\
	&\quad\quad c(\kappa)=\left(\frac{3}{g(\kappa)}-2\kappa^2+\frac{2-\kappa^2}{\sqrt{3}}k_{\text{max}}\right)\frac{1}{3 \, k^3_{\text{max}}}\\
	&\quad\quad g(\kappa)=\frac{3}{\left(1+\sigma \kappa^2\right)\kappa^2}\text{,}\quad\sigma\cong 0.596\;.
\end{align*}
This formula from \cite{KP2008} is remarkably accurate when compared to numerical curves; it should suffice for any application that needs an analytical form for the $k$-dependent growth rate of the gray soliton snake mode.

%\newpage

\end{document}